\documentclass[twocolumn,pre,superscriptaddress,preprintnumbers,amsmath,amssymb]{revtex4}

\usepackage{graphicx}
\usepackage{dcolumn}
\usepackage{bm}
\usepackage{color}

\newcommand{\gammadot}[0]{\dot{\gamma}}

\newcommand{\odif}[2]{\frac{\partial #1}{\partial #2}}

\newcommand{\ave}[1]{\langle {#1} \rangle}

\renewcommand{\phi}{\varphi}
\newcommand{\be}{\begin{equation}}
\newcommand{\ee}{\end{equation}}
\newcommand{\phig}{\varphi_{\rm G}}
\newcommand{\phij}{\varphi_{\rm J}}

\begin{document}

\title{Disentangling glass and jamming physics in the rheology of
soft materials}

\author{Atsushi Ikeda}
\affiliation{Laboratoire Charles Coulomb,
UMR 5221 CNRS and Universit\'e Montpellier 2, Montpellier, France}

\author{Ludovic Berthier}
\affiliation{Laboratoire Charles Coulomb,
UMR 5221 CNRS and Universit\'e Montpellier 2, Montpellier, France}

\author{Peter Sollich}
\affiliation{King's College London, Department of Mathematics, 
Strand, London WC2R 2LS, United Kingdom}

\date{\today}
\begin{abstract}
The shear rheology of soft particles systems becomes complex
at large density because crowding effects may induce 
a glass transition for Brownian particles, or a jamming transition
for non-Brownian systems. Here we successfully explore the hypothesis 
that the shear stress contributions from glass and jamming physics 
are `additive'. We show that the experimental 
flow curves measured in a large variety of soft materials 
(colloidal hard spheres, microgel suspensions, 
emulsions, aqueous foams) as well 
as numerical flow curves obtained for soft repulsive particles 
in both thermal and athermal limits are well described by
a simple model assuming that glass and jamming rheologies
contribute linearly to the shear stress, 
provided that the relevant scales for time and stress 
are correctly identified in both sectors. 
Our analysis confirms that the dynamics
of colloidal hard spheres is uniquely controlled by glass physics 
while aqueous foams are only sensitive to jamming effects.
We show that for micron-sized emulsions both contributions are 
needed to successfully account for the flow curves, which 
reveal distinct signatures of both phenomena. Finally, for two 
systems of soft microgel particles we show that the flow curves 
are representative of the glass transition of colloidal systems, 
and deduce that microgel particles are not well suited to 
studying the jamming transition experimentally.
\end{abstract}



\maketitle

\section{Introduction}

The emergence of solidity in disordered assemblies of soft 
particles is observed in large variety of systems, 
everyday examples including toothpaste, shaving foam, or 
paints~\cite{larson,coussot}.
Because of their practical use, a fundamental understanding of 
the flow property of dense suspensions is necessary, but this continues to 
represent a considerable challenge to physicists. In
particular, achieving a detailed understanding of fluid-to-solid 
transitions in amorphous materials is recognized as
an important challenge for modern condensed matter 
physics~\cite{rmp,jammingrev}. 

Two types of dense particulate systems have been considered.
One class is composed of particles that are small enough 
to be sensitive to thermal fluctuations. Brownian forces are 
for instance relevant in colloidal suspensions of micron-sized particles 
or smaller. In that case, colloids undergo collisions with 
the solvent molecules which play the role of a thermal bath.
The emergence of amorphous solids at large density in such materials 
is usually called {\it colloidal glass transition}~\cite{pusey},
by analogy with the molecular glass transition observed in 
organic or polymeric liquids~\cite{rmp}. 
  
In a second class of systems thermal fluctuations 
play a negligible role, with aqueous foams being one example.
In these non-Brownian
systems, the particle size is typically very large 
(say, in the millimeter range), such that the effect 
of collisions with solvent molecules is so weak that it
can be safely neglected on typical experimental timescales. 
The emergence of solidity in athermal amorphous systems 
was termed  {\it jamming transition}~\cite{liunagel}, by analogy with 
dry granular media~\cite{grains}.
 
Surprisingly, despite large differences in 
the underlying microscopic dynamics, the phenomenology of 
glass and jamming transitions appears remarkably similar, in particular 
when soft particle systems are considered~\cite{ohern1,vanhecke}.  
Below some critical density, the system can flow under infinitesimal 
shear stress. The nature of the flow is Newtonian, and the system
thus behaves as a viscous fluid. 
Above that critical density, the system does not flow if the shear stress
is below a characteristic finite value, called the yield stress.
The system is now a soft solid.  
Close to the critical density, the rheology is typically 
strongly non-linear: it is very sensitive to small density changes,
and the microscopic dynamics  becomes spatially and 
temporally heterogeneous~\cite{book}. 

This apparent similarity suggests that theoretical progress for one
type of systems might impact research for the other. For 
Brownian systems, flow curves are typically analyzed
by assuming that the glassy dynamics 
arising in thermal suspensions at rest is disrupted by 
the externally imposed shear flow~\cite{yamamoto,BBK,BB,fuchs}. 
A number of experimental 
studies, in particular in the colloidal literature, have then 
used these ideas to organize and quantitatively describe 
flow curves measured in the vicinity
of the glass transition~\cite{exp1}.
By contrast, much less is 
understood regarding the microscopic dynamics of athermal suspensions 
near the jamming transition, because the imposed shear flow
represents at the same time the external driving force 
and the source of microscopic fluctuations responsible for particle motion.
Nevertheless, detailed numerical analysis of model systems 
and theoretical modelling have suggested specific 
functional forms and scaling procedures to organize rheological 
data around the jamming 
transition~\cite{olson,tighe,olson2,lerner}. 
In fact these ideas have been used in a larger variety of systems, including 
atomistic glasses~\cite{egami} 
and colloidal particles~\cite{yodh}, where their relevance is not
obvious a priori.

Recently, we have used numerical simulations to study the rheology of dense 
assemblies of soft repulsive spheres~\cite{letter}. Despite its simplicity, 
this model is useful as it is known to display glassy dynamics 
at finite temperatures~\cite{tom}, and to undergo a geometric 
jamming transition at zero temperature~\cite{ohern1,durian}. 
By varying the relative 
strength of energy dissipation and thermal agitation over 
a broad range, we have demonstrated that glass and jamming 
physics impact the steady state flow curves over distinct 
stress scales and time scales, which can be varied independently. 
These results confirm that the two phenomena are actually distinct, 
and that their respective influences on rheological behaviour should
not be confused. 
In addition, we have suggested that the numerical flow curves 
are well described by a simple `additive' model, where 
the glass and jamming contributions to the shear stress can be captured
separately and then combined linearly into a unified model~\cite{letter}.

In the present work, we expand our presentation of this additive model
to explain in greater detail its construction and application 
to our numerical data. We then confront our rheological model 
deduced from the analysis of a simplistic soft particle system
to a broad range of experimental data taken from the literature. 
Overall, our analysis validates the hypothesis that,
despite their strongly nonlinear nature, glass and jamming physics 
linearly contribute to the rheology of soft particle systems
and act over distinct sectors, whose relative importance depends 
on the experimental system at hand. We also show that our 
model and theoretical analysis provide useful tools to organize 
and explain rheological data stemming from different 
experimental sources and materials. These tools can be used to 
efficiently identify which of the glass or jamming 
contributions is most relevant, or whether both types of 
physics are in fact needed to correctly describe the data. 

Our paper is organized as follows. In Sec.~\ref{construction} 
we construct our additive rheological model which we apply 
to numerical data obtained from simulations of harmonic spheres.
In Sec.~\ref{experiments} we use this model to analyze
steady state flow curves measured in colloidal hard spheres, 
microgel suspensions, emulsions, and aqueous foams.
In Sec.~\ref{conclusion}, we summarize our findings 
and discuss some perspectives for future work.

\section{`Additive' model for glass and jamming rheology} 

\label{construction}

\subsection{Flow curves for harmonic spheres} 

\label{harmonic}

To construct a general model for the complex rheology of 
dense suspensions, we use the numerical results obtained for 
a specific model as a reference. In previous work, 
we have studied the steady state rheology of harmonic spheres 
in the overdamped limit~\cite{letter}.
This can be seen as a simple model to describe the physical 
behavior of dense suspensions of deformable spheres, such as
the emulsions and colloidal suspensions considered in 
Sec.~\ref{experiments}.
 
Since the details of the model were already reported in our previous 
work, we only briefly summarize the key ingredients.
We consider $N$ harmonic spheres contained in 
a volume $V$. We use Lees-Edwards periodic boundary conditions~\cite{allen}
and solve the following 
equations of motion numerically:
\begin{eqnarray}
\xi (\odif{\vec{r}_i}{t} - \gammadot y_i \vec{e}_x) = - \sum_{j=1}^N  
\odif{v(|\vec{r}_i - \vec{r}_j|)}{\vec{r}_i} + \vec{R}_i.  \label{eom}
\end{eqnarray}
Here $\vec{r}_i$ represents the position of particle $i$, 
$y_i$ its $y$-component, and  $\vec{e}_x$ the unit vector 
along the $x$-axis. 
The damping coefficient, $\xi$, and the random force, $\vec{R}_i(t)$, obey 
the fluctuation dissipation relation:
$\ave{\vec{R}_{i,\alpha}(s) \vec{R}_{j,\beta}(s')} = 2 k_B T \xi \delta_{ij} 
\delta_{\alpha\beta} \delta(s-s')$, where 
$T$ is the temperature and $k_B$ is Boltzmann's constant. 
The interaction potential is a purely repulsive 
harmonic interaction truncated at the particle diameter, 
\begin{equation}
v(r) = \frac{\epsilon}{2} \left( 1 - \frac{r}{a} \right)^2 \, \Theta (r - a),
\end{equation} 
where $a$ is the particle diameter and $\Theta (x)$ is the 
Heaviside function. To avoid crystallization issues, 
we work with a 50:50 binary mixture of particles with diameter 
ratio 1.4, but this is otherwise largely irrelevant.  

By construction, the system has two characteristic energy scales, namely 
the thermal energy, $k_B T$, and the interaction energy of 
particles $\epsilon$. The ratio of these energy scales, 
$k_B T/\epsilon$, is an important control parameter.
Physically, it is a measure of the particle softness, 
expressed in units of kinetic energy. 

Glass and jamming effects can be distinguished most clearly in the
low-softness limit $k_B T/\epsilon \to 0$. This can be obtained 
in two ways. One can either 
take the limit of vanishing kinetic energy, $T \to 0$, at 
fixed repulsion energy scale $\epsilon$. One then studies the 
jamming transition 
of athermal packings of soft repulsive spheres. The alternative
is to keep temperature constant and send $\epsilon \to \infty$, 
a limiting case which corresponds to studying the physics of 
the thermalized hard sphere fluid. 

The two energy scales $k_BT$ and $\epsilon$ naturally provide two 
characteristic time scales and 
stress scales. 
The microscopic time scale for Brownian motion is 
\begin{equation}
\tau_{\rm T} = \frac{\xi a^2}{k_B T},
\label{taut}
\end{equation} 
and represents the time it takes a particle 
to diffuse by Brownian motion over a length scale comparable to its size.
The characteristic time scale 
for energy dissipation  
is given by 
\begin{equation}
\tau_0 = \frac{\xi a^2}{\epsilon} =   \frac{k_B T}{\epsilon} \tau_{\rm T} .
\label{tau0}
\end{equation}
  
Likewise, we can define two stress scales: a typical stress created 
by thermal fluctuations, 
\begin{equation}
\sigma_{\rm T} = \frac{k_B T}{a^3},
\label{sigmat}
\end{equation}
and an athermal stress scale:
\begin{equation}
\sigma_0 = \frac{\epsilon}{a^3} = \frac{k_B T}{\epsilon} \sigma_{\rm T} .
\label{sigma0}
\end{equation} 
The above expressions make it clear that these time scales and stress scales
are comparable when particle softness is large enough, ${k_B T} \approx 
{\epsilon}$, while they become increasingly separated as the athermal
limit is approached, ${k_B T} \ll {\epsilon}$.

In terms of the time and stress scales above we can now give a more
explicit statement of when the limit $k_B T/\epsilon\to 0$ corresponds
to athermal or thermal behaviour. If the limit is taken by lowering
$T$ at constant $\epsilon$ then the shear rate in athermal units,
$\dot\gamma \tau_0$, stays fixed: we are exploring the `athermal
sector'. Here we then also expect stresses to be on the athermal
scale, with $\sigma/\sigma_0$ of order unity. If on the other hand we
fix $T$ and take $\epsilon\to\infty$, then the shear rate on the
thermal scale, $\dot\gamma\tau_T$, remains finite. One then probes the
`thermal sector', with correspondingly
$\sigma/\sigma_T=O(1)$. Numerically, we find (see below) that the
prefactors in these estimates are such that, below and around the
glass transition, the interesting physics typically takes place for
$\dot\gamma\tau_T<1$ and $\sigma/\sigma_T<1$. As a matter of general
orientation we can therefore say that phenomena where the shear rate
and stress in thermal units are below unity belong to the thermal
sector, while the athermal sector corresponds to values of these
quantities significantly above one. 
A numerical analysis along the same lines 
has appeared recently~\cite{olson3}.

\begin{figure}
\includegraphics[width=\hsize]{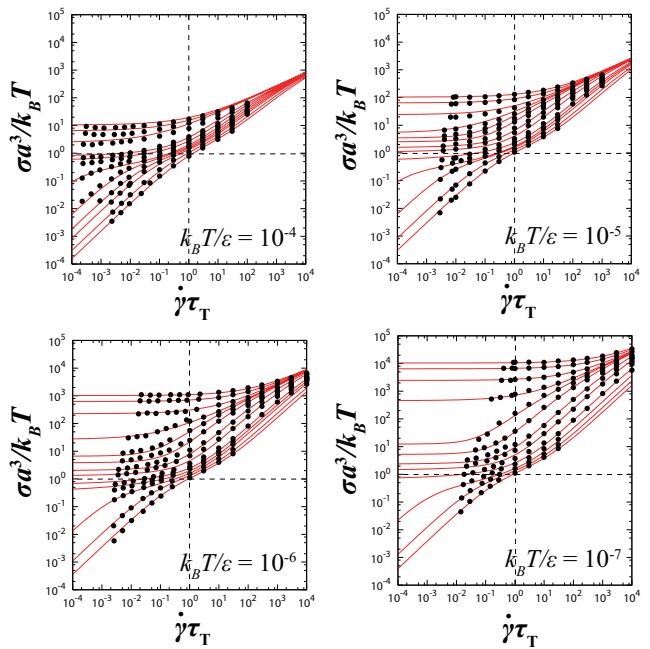} 
\caption{Flow curves of harmonic spheres for temperatures from 
$k_BT/\epsilon=10^{-4}$ down to $10^{-7}$ measured in the simulations
first described in Ref.~\cite{letter} (symbols). In each panel, packing 
fraction increases from bottom to top in the range 
$\varphi = 0.55 - 0.70$. The red lines are fits 
using the additive model presented in this work, with parameters
listed in Table~\ref{tab1}. 
The dashed lines delimit the thermal/glass (bottom left) 
and athermal/jamming (top right) sectors.}
\label{fig1}
\end{figure}

In Fig.~\ref{fig1} we present flow curves obtained by measuring 
the average shear stress, $\sigma$, under steady state conditions
created by a constant applied shear rate, $\gammadot$, and varying 
the packing fraction, $\varphi$, and the temperature, $T$, across 
a broad range.  Note that we use stresses
and shear rates normalized by $\tau_{\rm T}$ and $\sigma_{\rm T}$ as 
time and stress units; these are the appropriate scales 
for thermal systems. This leads us to introduce 
a `renormalized' shear rate, 
\begin{equation}
P_{\rm e} = \gammadot \tau_{\rm T},
\end{equation}
which is also called the P\'eclet number. 
Because the results shown in Fig.~\ref{fig1} were 
discussed in detail in Ref.~\cite{letter},
we only summarize the most salient features of these data. 

\begin{itemize}

\item {\it High temperature / Soft particles (such 
as $k_BT/\epsilon = 10^{-4}$).} 
A single fluid-to-solid transition is observed in this 
thermal regime. When the density 
is low, Newtonian flow is observed at low shear rates $\gammadot \ll 
\tau_{\rm T}^{-1}$. 
With increasing density, the Newtonian viscosity grows and a 
shear stress plateau appears at finite shear rate, 
corresponding to a strong shear-thinning regime. 
Near the transition density, $\varphi \approx \phig$,
the viscosity becomes very large, 
and a finite yield stress appears, which grows smoothly with density 
upon further compression. 

\item {\it Low temperature / Hard particles (such 
as $k_BT/\epsilon = 10^{-7}$).} 
The glass and jamming transition can be observed separately. 
When the density is low, two Newtonian regimes appear for $\gammadot \ll 
\tau_{\rm T}^{-1}$ and $\tau_{\rm T}^{-1} \ll \gammadot \ll \tau_0^{-1}$, 
respectively. 
The first Newtonian viscosity becomes very large 
near the glass transition density $\phig$ where 
the system acquires  
a finite yield stress, of the order of the thermal stress 
scale $\sigma_{\rm T}$.
However the second Newtonian 
viscosity remains finite at the glass transition density. 
On further increasing the density, the second Newtonian viscosity then
diverges at the jamming transition density $\phij$. 
At the same time, the yield stress value increases rapidly with
density; above the jamming transition it 
reaches a value which is controlled by the athermal stress 
scale $\sigma_0$.
\end{itemize}

These numerical results directly illustrate 
that glass and jamming transitions represent distinct ways for the system
to form an amorphous solid phase, since the fluid-to-solid transitions 
occur over different time windows (or, equivalently, different shear rates)
and give rise to solids with yield stress having different 
scales and different physical origins. By tuning the temperature 
of our system, we can move from Brownian soft particles 
undergoing a glass transition in the thermal sector 
of Fig.~\ref{fig1} to non-Brownian ones undergoing a 
geometric jamming transition in a different sector. Interestingly,
both types of physics compete and affect the flow curves 
for intermediate values of the temperatures. 
These observations indicate that the particle softness, $k_B T / \epsilon$, 
is a central control parameter for the rheology of soft materials. 
In particular, we note that when the particles are too soft
(equivalently, when temperature is too large), 
the jamming transition has no effect on the rheological data. 

\subsection{Construction of an additive model}

\subsubsection{The additive hypothesis  }

Using the simulation results as a reference, we now construct a 
simple rheological model. 
The major assumption is the additivity of the contributions from 
thermal and athermal parts to the total shear stress:  
\begin{eqnarray}
\sigma (T, \varphi, \dot{\gamma}) = \sigma_{\rm G} (T, \varphi, \dot{\gamma}) 
+ \sigma_{\rm J} (\varphi, \dot{\gamma}) + \eta_{\rm s} \dot{\gamma}.
\label{full} 
\end{eqnarray}
Here $\sigma_{\rm G}$ is the thermal part of the stress describing the 
physics of the glass transition, while $\sigma_{\rm J}$ is the athermal part 
of the stress accounting for the jamming transition. Finally,   
$\eta_s$ is the solvent viscosity,  so that $\eta_s \dot{\gamma}$ is 
the stress stemming from the background solvent. This only becomes 
relevant in the dilute limit but serves as a useful
reference value when considering experimental data.

Our goal with the model in Eq.~(\ref{full}) is not to make new types of 
predictions for the functional form of the flow curves  
for systems near glass and jamming transitions, but rather to explore
the interplay between both physics. Thus, we make use 
of previous theoretical work and choose simple 
functional forms that most efficiently capture the 
physical features of the flow curves 
associated with each of these transitions. 

\subsubsection{The glass contribution }

First, we specify 
the thermal part of the stress, $\sigma_{\rm G}$. We write it as the sum 
of the zero and the finite shear rate parts: 
\begin{eqnarray}
\frac{\sigma_{\rm G} (T, \varphi, \dot{\gamma})}{ \sigma_{\rm T} } &=& 
\sigma_{\rm GY} (\varphi) + \label{sgmg} \\
&& \frac{Y_{\rm G}}{(\dot{\gamma} \tau_{\rm T} G(\varphi))^{-1} + 
(1+p_{\rm G}(\dot{\gamma} \tau_{\rm T})^{\alpha_{\rm G}})^{-1}} \nonumber. 
\end{eqnarray}
Although compact, this expression incorporates quite a number 
of physical features of the rheology of glass-forming 
suspensions. In this equation, $Y_{\rm G}$ and $p_{\rm G}$ are 
two dimensionless numerical prefactors, with no deep 
information content. This expression is constructed 
to produce flow curves with no yield stress 
(defined as the $\gammadot \to 0$ limit value of 
the shear stress) at low density, and a finite 
value for the glass phase when 
density is larger than the glass transition 
density, $\varphi_{\rm G}$, which is assumed to be sharply
defined. This last point is discussed further in Sections~\ref{fitting} 
and~\ref{beyond_mct} below.

We note to start with that the shear rate and 
shear stress in Eq.~(\ref{sgmg}) are expressed in units 
appropriate for Brownian suspensions, namely
$\sigma_{\rm T}$ and $\tau_{\rm T}$.
It is mainly via these elementary
units $\tau_{\rm T}$ and $\sigma_{\rm T}$ 
that the temperature dependence of the flow curves enters
(apart from a minor temperature dependence of the transition 
density, see Sec.~\ref{fitting} below). 
The microscopic expressions of these units are given by 
Eqs.~(\ref{taut}, \ref{sigmat}) for a system obeying Langevin dynamics,
as in our numerical simulations.
In experiments, one might want to replace the 
damping coefficient $\xi$ in this expression using instead 
Stokes' law to obtain the microscopic time scale: 
\begin{equation}
\tau_{\rm T} = \frac{3 \pi \eta_{\rm s} a^3}{k_B T}.
\label{stokes}
\end{equation} 

Having set the correct scales, we now describe the 
functional form of the flow curves predicted by Eq.~(\ref{sgmg}). 
For densities below the glass transition, $\varphi < 
\varphi_{\rm G}$, the yield stress is zero, and we impose 
accordingly $\sigma_{\rm GY}(\varphi \leq \varphi_{\rm G}) = 0$.
Thus, only the second term on the right hand 
side needs to be discussed. For small 
shear rates,  $\dot{\gamma} \ll (\tau_{\rm T} 
G(\varphi))^{-1}$, 
the first term in the denominator dominates and 
the equation becomes $\sigma_{\rm G} \approx Y_{\rm G} \sigma_{\rm T} 
\tau_{\rm T} G(\varphi) \dot{\gamma}$. 
This represents Newtonian behavior with the {\it thermal viscosity} 
$\eta_{\rm T} (\phi)$ given by  
\begin{equation}
\eta_{\rm T}(\varphi) \equiv 3\pi Y_{\rm G} G(\varphi) \eta_s.
\label{thermal_visc}
\end{equation}
This expression shows that the dimensionless function $G(\varphi)$
controls the rapid growth of the viscosity on approaching
the glass transition, and thus also corresponds to  
the growing equilibrium relaxation time of the unsheared Brownian 
suspension at density $\varphi$, via
the relation $\tau_\alpha \approx G(\varphi) \tau_{\rm T}$.  

At larger shear rates, the flow curve (\ref{sgmg}) becomes $\sigma_{\rm G} 
\approx Y_{\rm G} \sigma_{\rm T} (1 + p_{\rm G}(\dot{\gamma} 
\tau_{\rm T})^{\alpha_{\rm G}})$; here we assume that the exponent
$\alpha_{\rm G}$ takes a value below unity. 
This expression describes the strong shear-thinning regime 
obtained in dense fluids when the shear rate competes 
with the slow glassy dynamics and drives the system away from equilibrium. 
This competition is modelled by a plateau regime, $\sigma_{\rm G} \approx Y_G 
\sigma_{\rm T}$, when $\dot{\gamma} \ll \tau_{\rm T}^{-1}$, 
finally followed by a different shear-thinning behaviour, 
$\sigma_{\rm G} \approx Y_G p_{\rm G} \sigma_{\rm T} (\dot{\gamma} 
\tau_{\rm T})^{\alpha_{\rm G}}$. These two shear-thinning 
regimes result from the well-known existence of two 
separated relaxation processes ($\alpha$ and $\beta$ relaxations) 
in highly viscous fluids~\cite{thomas}. 

In the glass phase, $\varphi > \varphi_G$, we impose 
an infinite value of the Newtonian viscosity using $G(\varphi)=\infty$, 
such that the first term in the denominator disappears from 
Eq.~(\ref{sgmg}) and
the flow curve simplifies to 
$\sigma_{\rm G} / \sigma_{\rm T} = [Y_{\rm G} + \sigma_{\rm GY}(\varphi)] + 
Y_{\rm G} p_{\rm G} (\dot{\gamma} \tau_{\rm T})^{\alpha_{\rm G}}$. 
This describes the existence of a finite yield stress, 
\begin{equation}
\sigma^{\rm yield} =  
\left[ Y_{\rm G} + \sigma_{\rm GY}(\varphi) \right] \sigma_{\rm T}  , 
\label{yieldglass}
\end{equation}
whose scale is set by the thermal stress. At finite shear rates the
rheology is of Herschel-Bulkley type~\cite{larson}, 
with shear-thinning exponent $\alpha_{\rm G}$. 

\begin{figure}
\includegraphics[width=\hsize]{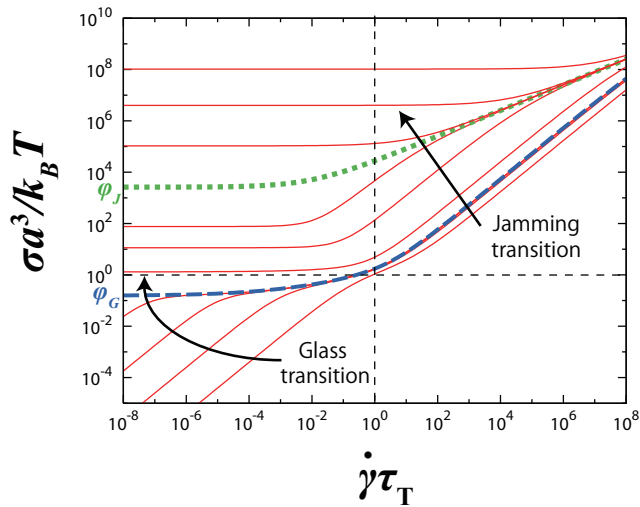} 
\caption{Flow curves predicted by the additive model 
at low temperatures, $k_BT/\epsilon=10^{-10}$,   
when glass (bottom left) and jamming (top right) sectors are 
well separated. Notice the different functional forms of the 
flow curves near the critical densities $\phig$ and $\phij$ 
(highlighted with dashed and dotted lines), reflecting the 
distinct microscopic dynamics associated with thermal 
and athermal situations.}
\label{fig3}
\end{figure}

The resulting functional forms for the glass contribution 
are shown graphically in the bottom left sector of Fig.~\ref{fig3}, which 
corresponds to the thermal sector of the rheology predicted 
by the present model. We defer a discussion of the volume fraction dependences
of $\sigma_{\rm GY}$ and $G$ to Sec.~\ref{fitting} below.

\subsubsection{The jamming contribution }

We now turn to the jamming contribution to the shear stress in 
Eq.~(\ref{full}),
$\sigma_{\rm J}$.
We express this athermal part of the stress as: 
\begin{equation}
\frac{ \sigma_{\rm J} (\varphi, \dot{\gamma}) }{\sigma_0 } =
\sigma_{\rm JY} (\varphi) + \frac{Y_{\rm J}}{(\dot{\gamma} 
\tau_0 J(\varphi))^{-1} + (p_{\rm J}(\dot{\gamma} 
\tau_0)^{\alpha_{\rm J}})^{-1}}
\label{sgmj} 
\end{equation}
In this expression, we have introduced two dimensionless 
parameters, $Y_{\rm J}$ and $p_{\rm J}$. 
In contrast to the glass contribution, the stress and time units
are now $\sigma_0$ and $\tau_0$,
whose microscopic expressions were given in Eqs.~(\ref{tau0}, 
\ref{sigma0}). Therefore, temperature does not enter the stress
contribution~(\ref{sgmj}),
as expected from the athermal nature of jamming physics. 

Expression~(\ref{sgmj}) has many similarities with Eq.~(\ref{sgmg}) but is 
qualitatively slightly simpler, as we now describe.
For densities below the jamming density, we impose 
$\sigma_{\rm JY}(\varphi \leq \varphi_{\rm J}) = 0$, such that 
a Newtonian viscosity emerges in the low shear rate
limit, $\dot{\gamma} \ll (\tau_0 J(\varphi))^{-1}$.
Here $\sigma_{\rm J} \approx Y_{\rm J} 
\sigma_0 \tau_0 J(\varphi)$, which defines 
the {\it athermal viscosity} 
\begin{equation}
\eta_0 (\varphi) \equiv 3\pi Y_{\rm J} 
J(\varphi) \eta_s.
\end{equation}
This equation shows that the dimensionless function 
$J(\varphi)$ now controls the divergence of the Newtonian 
viscosity on appraoching $\varphi_{\rm J}$.  The difference 
with the glass regime is seen at larger shear rates because
the constant contribution in the denominator of Eq.~(\ref{sgmg}) is absent 
in Eq.~(\ref{sgmj}).
As a result, Eq.~(\ref{sgmj}) shows a simple power law shear-thinning 
behavior 
$\sigma_{\rm J} \approx Y_{\rm J} \sigma_0 p_{\rm J}(\dot{\gamma} 
\tau_0)^{\alpha_{\rm J}}$, instead of the plateau in Eq.~(\ref{sgmg}).
This reflects the absence of distinct $\alpha$ and $\beta$ relaxations
in athermal systems~\cite{letter}. 

In the jammed phase, $\varphi > \varphi_{\rm J}$, 
we impose a nonzero yield stress, 
$\sigma_{\rm JY} > 0$, 
and an infinite viscosity via $J(\varphi) = \infty$. As a result, 
we obtain $\sigma_{\rm J}/\sigma_0 = \sigma_{\rm JY}(\varphi) + Y_{\rm J} p_J 
(\gammadot \tau_0)^{\alpha_{\rm J}}$, which 
describes the existence of a finite yield stress, 
\begin{equation}
\sigma^{\rm yield} =  \sigma_{\rm JY} (\varphi) \, \sigma_0.
\label{yieldjam}
\end{equation} 
Its units are given by the athermal stress. Again at finite shear
rates we have Herschel-Bulkley rheology, with the 
shear-thinning exponent now $\alpha_{\rm J}$. 

The resulting functional forms for the jamming contribution 
are shown graphically in the top right quadrant of Fig.~\ref{fig3}, which 
corresponds to the athermal sector of the rheology predicted 
by the present model. 
 
\subsection{Density dependence: Choice of fitting functions}

\label{fitting}

In both thermal and athermal expressions for the shear stress, 
we have described a change from fluid to solid behaviour signalled
by rapidly growing viscosities controlled by the dimensionless 
functions $G(\varphi)$ and $J(\varphi)$, 
and emerging yield stresses controlled by 
the functions $\sigma_{\rm GY}(\varphi)$ and $\sigma_{\rm JY}(\varphi)$.
Within our model, 
these functions entirely control the density dependence of the flow
curves. In this subsection, we describe the physics 
embodied by these quantities, and motivate the specific choices 
we made to fit numerical and experimental flow curves in the present paper. 

\subsubsection{Diverging Newtonian viscosities }

For thermal systems, the function $G(\varphi)$ controlling the 
growth of the Newtonian viscosity has been carefully analyzed in a large 
number of studies, as it quantifies the dynamic slowing down 
of a hard sphere fluid on its approach to the colloidal 
glass transition~\cite{pusey,chaikin,bill,gio}. 
As is well known from the literature of glass-forming materials~\cite{rmp}, 
this timescale growth cannot be represented by a single functional form 
because the slowing down of the dynamics crosses over
from power law timescale increase at moderate volume fraction, to a steeper 
`activated' exponential growth at larger density~\cite{gio}. This 
crossover can be understood theoretically as a change of 
relaxation mechanism from collective but nonactivated
dynamics described by mode-coupling theory~\cite{gotze}, 
to correlated activated events closer to the glass transition~\cite{rmp}.  

In practice, however, given the limited range of viscosity 
data that is accessible in typical numerical and 
experimental work, a single functional form is often a good 
enough approximation to describe the data. 
Therefore, in the following we shall describe the data 
using a power law divergence: 
\begin{equation}
G(\varphi) = h_{\rm G} (\phig - \varphi)^{-\gamma_{\rm G}}, 
\label{gphi}
\end{equation}
where $h_{\rm G}$ is a numerical prefactor. The location of
the glass transition, $\varphi_{\rm G}$, and the associated 
exponent, $\gamma_{\rm G}$, are fit parameters that serve to describe 
the increase with density of the thermal viscosity $\eta_{\rm T}(\varphi)$. 
As is clear from the above description, the value 
extracted for the location of the glass transition has the same physical
content as the location of the mode-coupling temperature
for standard glass-forming materials; it should not be confused 
with the location of a genuine dynamic or thermodynamic 
singularity~\cite{tom2}.  
 
On the other hand, much less is known about the 
density dependence of the athermal viscosity $\eta_0$, controlled 
in our model by the dimensionless function $J(\varphi)$. 
Theoretical~\cite{olson,lerner} and experimental~\cite{lindner,boyer} 
studies seem to indicate 
that a power law divergence describes the results well, 
and accordingly we use
\begin{equation}
J(\varphi) = h_{\rm J} (\phij - \varphi)^{-\gamma_{\rm J}}, 
\label{jphi}
\end{equation}
with $h_{\rm J}$ a numerical prefactor. The exponent
$\gamma_{\rm J}$ is found numerically and experimentally to 
be close to $\gamma_{\rm J} = 2$. 
The critical density, 
$\phij$, appearing in the expression (\ref{jphi}) represents the 
location of the jamming transition.

Although Eqs.~(\ref{gphi}, \ref{jphi}) appear similar, 
we note that the former has been derived from 
a microscopic perspective~\cite{gotze}, 
while the second one is only supported 
by experimental data or detailed theoretical and numerical 
analysis performed over a limited range of densities~\cite{lerner,boyer}. 
Therefore, 
while empirically robust, the status of the algebraic divergence
for non-Brownian hard sphere suspensions remains to be clarified. 
A second difference between the two expressions concerns the nature  
of the critical densities. While we described 
$\varphi_{\rm G}$ as the crossover density emerging from 
a mode-coupling theory analysis of the data, an actual divergence
is expected in the athermal limit, such that $\varphi_{\rm J}$ is 
not simply a crossover but a genuine critical density~\cite{jammingrev}. 
However, its location remains ambiguous: the nonequilibrium nature 
of the jamming transition implies that it
depends on the specific details of the procedure used to bring a
system close to jamming~\cite{donev}. This implies, for instance, that
the values for $\phij$ obtained by slow or fast compressions of random 
sphere packings do not coincide with the critical density 
measured under shear~\cite{pinaki}. Thus, $\phij$ in Eq.~(\ref{jphi}) is very 
specifically defined as the critical density that controls the 
divergence of the Newtonian viscosity of non-Brownian particles 
in the hard sphere limit~\cite{olson,lerner}.

\subsubsection{Emerging yield stress }

\begin{figure}
\includegraphics[width=\hsize]{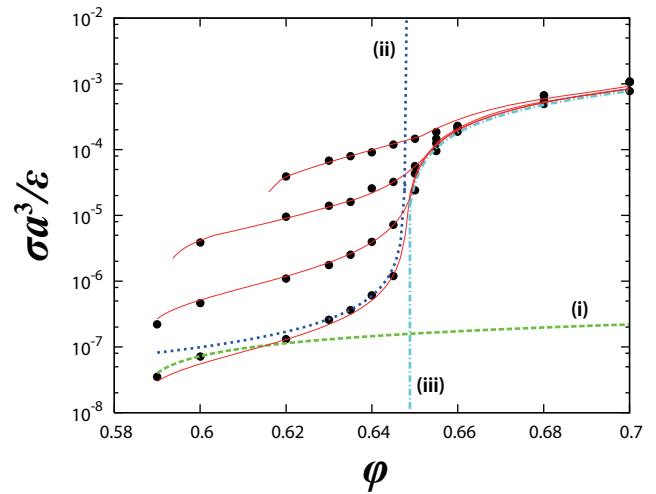} 
\caption{Yield stress of harmonic 
spheres, measured numerically at $k_BT/\epsilon=10^{-4}$, 
$10^{-5}$, $10^{-6}$ and $10^{-7}$ (from top to bottom).
These results are extracted from the $\gammadot \to 0$ limit 
of the flow curves in Fig.~\ref{fig1}. 
The red lines are fits using Eqs.~(\ref{yg}, \ref{yj}). 
The three lines for $T=10^{-7}$ represent fits to the simpler functional forms 
in Eq.~(\ref{ys1}) [(i), dashed],  Eq.~(\ref{ys2}) [(ii), dash-dotted], 
and Eq.~(\ref{ys3}) [(iii), dotted]. }
\label{fig2}
\end{figure}

We now discuss the behaviour of the yield stress for glass and jammed 
phases, which is also characterized by a number of singular behaviours, 
as demonstrated by the numerical measurements performed on harmonic spheres
at various temperatures and reported in Fig.~\ref{fig2}. 
These yield stress data 
are obtained from the zero shear rate extrapolation 
of the flow curves shown in Fig.~\ref{fig1}.

As can be seen from Fig.~\ref{fig2}, the yield stress data show three
distinct characteristic behaviours, which are more or less pronounced 
depending on temperature, and thus one can use various levels of 
sophistication in the description of these data.  

The first observation is related to the appearance and growth 
with density 
of a yield stress above the glass transition $\varphi_{\rm G}$. 
Within our model, this behaviour is controlled 
by the term $\sigma_{\rm GY} (\varphi)$, see Eq.~(\ref{yieldglass}). 
A possible functional form is 
\begin{equation}
\sigma_{\rm GY} (\varphi \geq \phig) = 
A_{\rm G} (\varphi - \phig)^{\beta_{\rm G}},  
\label{ys1} 
\end{equation}
where $A_{\rm G}$ is a prefactor and $\beta_{\rm G}$ an exponent 
which characterizes the growth of the yield stress with density.  
This form is consistent with the description 
provided by the extension of the mode-coupling 
theory to sheared suspensions~\cite{fuchs}, 
where the exponent is given by 
$\beta_{\rm G} = \frac{1}{2}$. The data 
in Fig.~\ref{fig2} indicate that Eq.~(\ref{ys1}) 
is sufficient to describe the yield stress of harmonic spheres 
when the temperature is high enough, e.g.\ $k_B T = 10^{-4}$, so that the
jamming transition does not play any role.  
Note, however, that a precise determination of the exponent 
in Eq.~(\ref{ys1}) is not possible, as it is difficult 
to obtain unambiguous yield stress data very near $\phig$ due to
the crossover nature of the mode-coupling transition:
the flow curve at the fitted $\phig$ would 
eventually become Newtonian at low enough shear rates.   

By contrast, the jamming density becomes relevant when temperatures 
becomes much smaller, and this affects the yield stress on both sides
of the jamming transition. For $\varphi < \varphi_{\rm J}$ and 
$T \to 0$ the behaviour of hard sphere glasses is recovered, 
for which the yield stress diverges as 
$\varphi_{\rm J}$ is approached from below~\cite{bookmewis}. 
One can model this behaviour using again an algebraic dependence, 
\begin{equation}
\sigma_{\rm GY}(\varphi \leq \phij) 
= B_{\rm G} (\phij - \varphi)^{-\beta_{\rm GJ}}, \label{ys2}
\end{equation}
where $B_{\rm G}$ is a numerical prefactor.
We are not aware of specific theoretical predictions for the value of
the exponent $\beta_{\rm GJ}$ governing the yield stress divergence, 
but one might expect that it is the 
same exponent which also controls the divergence of the 
pressure in compressed hard spheres, for which the value 
$\beta_{\rm GJ} = 1$ is well documented~\cite{tom2,freevol,wood}. 

The third characteristic density dependence is obtained 
above the jamming transition in the $T \to 0$ limit. In our model this
is entirely controlled by the term $\sigma_{\rm JY}$; 
see Eq.~(\ref{yieldjam}). The emergence of a yield stress in 
athermal soft sphere packings has been described in previous numerical
work as a continuous power law growth~\cite{durian,olson}, 
\begin{eqnarray}
\sigma_{\rm JY} (\varphi \geq \phij)
= Y_{\rm J} (\varphi - \phij)^{\beta_{\rm J}}. \label{ys3}
\end{eqnarray}
Note that there is no need to introduce a new prefactor in this 
expression, as the second term of Eq.~(\ref{sgmj}) already contains
two adjustable prefactors in the denominator, $p_{\rm J}$ and $h_{\rm J}$ 
from $J(\phi)$. 
The expression (\ref{ys3}) 
is known to be sufficient when describing fully athermal 
assemblies of soft particles~\cite{olson}. 
The exponent $\beta_{\rm J}$ has been found to be close 
to $\beta_{\rm J} = 1$, 
although small deviations from this simple value of $\beta_{\rm J}$ 
have also been discussed~\cite{olson}. 

We have represented the three fitting functions in  
Eqs.~(\ref{ys1}, \ref{ys2}, \ref{ys3}) as lines going through the 
numerical results in Fig.~\ref{fig2}, demonstrating that 
each of these expressions can describe a limited range of densities
fairly well; each therefore correctly captures a different physical regime  
of the yield stress behaviour of harmonic spheres. 

However, as is clear from Fig.~\ref{fig2}, for finite but very 
low temperatures all three regimes affect the density 
dependence of the yield stress. Therefore, to  
describe the functional form of such low-$T$ data correctly one must introduce 
more complicated model equations that interpolate smoothly 
between the various regimes. A possible expression is 
as follows: 
\begin{eqnarray}
&& \sigma_{\rm GY}  = 
\frac{1}{ Y_{\rm GJ}^{-1} \sqrt{ \frac{\sigma_T}{\sigma_0}} + 
Y_{\rm G}'^{-1} (\varphi - \phig)^{-\beta_{\rm G}} 
(\phij - \varphi)^{\beta_{\rm GJ}}}, \quad \label{yg} \\
&& \sigma_{\rm JY} =  Y_{\rm GJ} 
\sqrt{ \frac{\sigma_{\rm T}}{\sigma_0}} 
+ Y_{\rm J} (\varphi - \phij)^{\beta_{\rm J}}.   \label{yj}  
\end{eqnarray}
In these equations we have used the product 
$Y_{\rm G}' (\varphi - \phig)^{\beta_{\rm G}}(\phij - 
\varphi)^{-\beta_{\rm GJ}}$ in the thermal part to 
interpolate continuously between expressions 
(\ref{ys1}) and (\ref{ys2}) in between the glass and 
jamming transition densities. This has the added benefit of reducing
the number of adjustable numerical prefactors by one, by effectively replacing
$A_{\rm G}$ and $B_{\rm G}$ with $Y_{\rm G}'$.
On the other hand, we have also introduced 
the constant $Y_{\rm GJ}$ in order to extend 
the $T=0$ singularities in Eqs.~(\ref{ys2}) and (\ref{ys3}) to nonzero
temperatures while connecting them smoothly. Indeed, one verifies easily that 
exactly at $\phij$ the expressions (\ref{yg},\ref{yj}) give the same result
$\sigma^{\rm yield} = \sigma_{\rm GY} \sigma_{\rm T} 
= \sigma_{\rm JY} \sigma_{0} = Y_{\rm GJ} \sqrt{ \sigma_{\rm T} \sigma_0}$.   
Therefore, the dimensionless parameter $k_BT/\epsilon$ is again the 
key parameter 
controlling the typical stress scale at the crossover 
between thermal and athermal regimes, via $\sqrt{\sigma_{\rm T}/\sigma_0} = 
\sqrt{k_BT/\epsilon}$.
Note that the complex pattern of these yield stress data near jamming 
resembles the behaviour found for the pressure in thermalized 
packings of soft spheres near the jamming transition~\cite{hugo2}.   

For definiteness we note that in writing Eqs.~(\ref{yg}, \ref{yj}), 
we have assumed that $\sigma_{\rm JY}$ is zero for $\phi<\phij$ while 
$\sigma_{\rm GY}$ vanishes for $\phi>\phij$. The sudden drop of 
$\sigma_{\rm GY}$ to zero at $\phij$ does not, of course, have any 
physical meaning; only the sum $\sigma_T\sigma_{\rm GY} + \sigma_0 
\sigma_{\rm JY}$ matters in our model and this is a smooth function 
of the volume fraction $\varphi$. 

In Fig.~\ref{fig2}, the full lines through the numerical data 
are obtained by using simultaneously Eqs.~(\ref{yg}, \ref{yj}), which 
clearly gives very satisfying results. 
The fitting parameters used to describe the simulations 
are summarized in Table~\ref{tab1} below. We emphasize that 
the apparent complexity of these expressions and the relatively large
number of adjustable parameters are needed because we want to capture 
in a single set of equations an unusually large number of 
physical phenomena pertaining to the physics of both glass and jamming 
transitions, for both thermal and athermal systems, soft and hard 
particles.  

\subsection{The additive model at work}

Having carefully analyzed the behaviour of the yield stress in 
the previous section, we are now in a position to combine 
all the elements of the additive model to describe the 
flow curves obtained numerically for harmonic spheres at various 
temperatures and densities. 

To this end, we add the glass and jamming 
contributions, Eqs.~(\ref{sgmg}, \ref{sgmj}) respectively, to the
shear stress~(\ref{full}); the final solvent contribution can be
safely neglected at the volume fractions of interest.
Into these expressions we insert the fitting forms for
the viscosity divergences, Eqs.~(\ref{gphi}, \ref{jphi}), and the emerging 
yield stresses, Eqs.~(\ref{yg}, \ref{yj}). The details 
of the chosen parameter values are given in Table~\ref{tab1} below.   

This procedure yields the flow curves shown as continuous lines 
in Fig.~\ref{fig1}. In practice, we 
first tuned the parameters to reproduce the simulation results 
at $k_BT/\epsilon = 10^{-6}$. We then used these parameters 
to fit all other temperatures. 
The figure demonstrates that the complex behaviour observed in both 
glass and jamming limits, as well as the crossover between the 
two regimes, are well described by the additive model. This
supports the validity of our analysis in terms of 
a linear combination of independent contributions
respectively stemming from glass and jamming physics.

\subsection{Beyond the simplified mode-coupling description}
\label{beyond_mct}

We have commented already several times on the fact that our additive
model uses a simplified representation of the glass transition and its
effect on shear rheology, which is in the spirit of mode coupling
theory. Before proceeding to the analysis of experimental data, we
want to clarify that this simplification is not conceptually required
by the additive model, and could be removed if desired.

At first sight our way of writing the glass contribution to the
stress, Eq.~(\ref{sgmg}), seems to be tied to the existence of a
specific volume fraction $\phig$ above which the yield stress
$\sigma_{\rm GY} (\varphi)$ becomes nonzero. However, this
contribution can be written in an equivalent form as
\begin{eqnarray}
\frac{\sigma_{\rm G} (T, \varphi, \dot{\gamma})}{ \sigma_{\rm T} } &=& 
\frac{\sigma_{\rm GY} (\varphi) + Y_{\rm G}}{(\dot{\gamma} \tau_{\rm T} 
G(\varphi))^{-1} + 
(1+p_{\rm G}(\varphi)(\dot{\gamma} \tau_{\rm T})^{\alpha_{\rm
    G}})^{-1}}.
\label{sgmg2}
\end{eqnarray}
Here we have allowed a density-dependence of the numerical coefficient
$p_{\rm G}(\varphi)$. If this is chosen as $p_{\rm G}(\varphi)=p_{\rm
  G} Y_{\rm G}/(Y_{\rm G}+\sigma_{\rm GY}(\varphi))$, it is a simple
matter to check that (\ref{sgmg}) and (\ref{sgmg2}) are
identical. This is because for $\varphi<\phig$ we have $\sigma_{\rm
  GY}(\varphi)=0$, while for $\varphi>\phig$, $G(\phi)$ is infinite
and so the first term in the denominator vanishes.

The form (\ref{sgmg2}) of the glass stress is now easy to generalize
to more sophisticated representations of the behaviour of the glass
transition. One could represent the crossover to activated relaxation
processes by keeping $G(\varphi)$ finite but making it cross over to
an exponentially fast increase above $\phig$. In the same vein,
$\sigma_{\rm GY}(\varphi)$ could be given a smooth, non-singular onset
around $\phig$, and one could also allow a more general density
dependence for $p_{\rm G}(\varphi)$. Conceptually, these generalizations
are important: if $G(\varphi)$ remains finite, the model never predicts
a genuine yield stress for $\phi<\phij$ and there is always a finite
Newtonian viscosity, given by (\ref{thermal_visc}).
A final possibility would be to incorporate a genuine glass divergence 
located at the `ideal' glass transition density~\cite{rmp,tom2}.  

For representing real flow curves, on the other hand, it is
essentially irrelevant whether $G(\varphi)$ has a true divergence or
not. In the latter case, the value $G(\varphi)$ will still become so large
near $\phig$ that $\dot{\gamma} \gg (\tau_{\rm T} G(\varphi))^{-1}$
for all accessible shear rates. The predicted flow curve then exhibits
an effective yield stress, given by (\ref{yieldglass}) as before. Our
simpler representation of $G(\varphi)$ avoids introducing additional
parameters for the divergence above $\phig$, which cannot be
determined with any accuracy from the available 
rheological data and would not
improve the quality of our fits.

\section{Analysis of experimental results} 

\label{experiments}

\subsection{Overview} 

In this section, we will apply the set of equations described in 
Sec.~\ref{construction} to analyze experimental data stemming 
from a number of different sources. Our main goal is to demonstrate 
how to disentangle glass and jamming rheologies in experimental work, 
and detect whether there exist experimental systems where the crossover 
between both sectors revealed by our numerical work is 
relevant. The study of such systems would be useful in assessing 
the validity of the key aspect of the model---the additivity
hypothesis. Another expected outcome is a clearer understanding and 
classification of the nature of the formation of amorphous
solids in soft materials. 

For all systems considered, we shall need to determine 
the particle softness, the particle size and the solvent viscosity, 
which will serve to construct the elementary time and stress scales. 
In our additive model, the particle softness $k_BT/\epsilon$ is the key 
parameter governing the structure of the flow curves. 
As mentioned above, very soft particles will not be very sensitive 
to the jamming singularity. This is because the latter requires that the 
notion of particle 
contact is physically meaningful, which is not the case when large thermal
fluctuations are present. At low temperatures, on the other hand,
the two transitions 
are well-resolved and belong to distinct stress and time 
sectors as 
demonstrated by the flow curves in Fig.~\ref{fig3}.
For fixed low particle softness, the role of the particle size and 
solvent viscosity is thus to determine 
which sector of the flow curves is actually observed in any given experiment.

Before going into the details of the various systems, it 
is instructive to consider some numbers. 
In a typical soft matter 
experimental setup~\cite{larson}, measurable shear stress and shear rate 
scales are $\sigma \approx 1 \, {\rm Pa}$  
and $\gammadot \approx 1 \, {\rm s}^{-1}$. 
For a particle diameter of the order of 
$a = 0.1 \, \mu{\rm m}$ and solvent viscosity $\eta_s =
1 \, {\rm mPa \cdot s}$ (this is the typical value for water),  
one obtains $\sigma/\sigma_{\rm T} = \sigma a^3/k_BT \approx 1$  
and $\gammadot \tau_{\rm T} \sim 10^{-3}$. This implies that 
the thermal sector will be explored experimentally, and
glass transition physics should be dominant.
 
On the other hand, when the particle diameter is $a = 10 \, \mu{\rm m}$ with 
the same solvent, otherwise identical experimental conditions will 
correspond to $\sigma a^3/k_BT \approx 10^6$  and $\gammadot 
\tau_{\rm T} \sim 10^3$. Such an experiment would probe the athermal
sector of the flow curves, where a jamming transition can be observed.

This little exercise shows that an interesting interplay between 
glass and jamming physics can be observed only in an experiment 
with particle sizes of the order of $a \approx 1 \, \mu{\rm m}$, 
and for particles which 
are not too soft. We shall see below that emulsions 
seem to represent the best compromise. 
In the next subsections, we follow the above approach: we shall 
redraw several available experimental flow curves using, for
definiteness, units  
appropriate for thermal systems; we then fit these flow curves using
the model equations described in Sec.~\ref{construction}.

\subsection{Glass physics: PMMA colloids}

\label{pmma}

\begin{figure}
\includegraphics[width=\hsize]{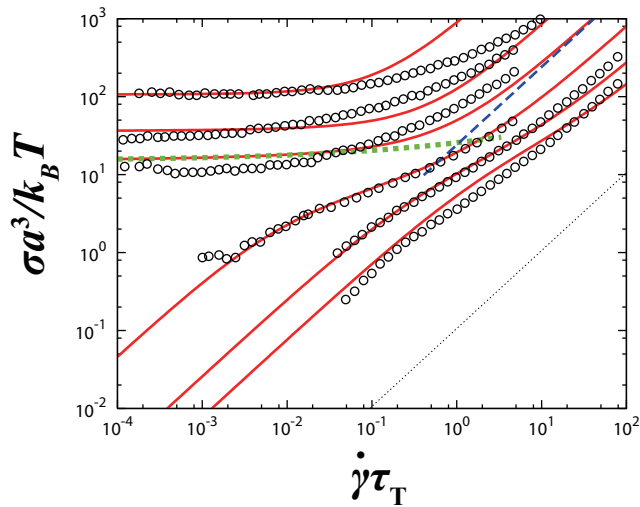} 
\caption{Experimental flow curves of a nearly hard sphere 
PMMA colloidal suspension, taken from Petekidis {\it et al.}~\cite{pete},
are shown using thermal units.   
The full lines are the fits using our additive model using  
$k_BT/\epsilon = 10^{-8}$ and $\phig = 0.575$; see also Table~\ref{tab1}.
For the first glass flow curve, we show the decomposition 
of the stress into its glass (dotted) and jamming (dashed) contributions.
The thin dashed line represents the dilute limit. 
Clearly, the fluid to solid transition is a colloidal
glass transition, with little influence from the jamming physics.}
\label{fig4}
\end{figure}

In this subsection, we analyze the flow curves reported 
in Ref.~\cite{pete} for nearly hard sphere colloids. 
This PMMA colloidal suspension is a very important experimental system 
whose rheology has been studied in much detail. This is because 
PMMA colloids are considered as a very good experimental 
realization of the hard sphere fluid~\cite{pusey}, 
which is itself an important model 
system for the statistical mechanics of simple fluids~\cite{hansen}. 
In particular, PMMA colloids have been used extensively in 
experimental studies of  
the hard sphere glass transition~\cite{pusey,chaikin,bill,gio,weeks}. 
In the specific study in Ref.~\cite{pete}, 
the particle diameter is $a = 0.36 \, \mu{\rm m}$ with 
a size polydispersity of about 12~\% to prevent crystallization. 
We can thus estimate the thermal stress scale as $\sigma_{\rm T}=
k_B T/a^3 = 0.0825 \, {\rm Pa}$. 
Since the free diffusion constant $D_0$ of this colloid is also 
reported in Ref.~\cite{pete}, 
the thermal time scale for this system can be estimated by 
$\tau_{\rm T} = a^2/D_0$ without using the Stokes' law Eq.~(\ref{stokes}). 
The estimated value is $\tau_{\rm T} = 0.158 \, {\rm s}$. 
We show the measured flow curves using these thermal units 
in Fig.~\ref{fig4}. This representation quickly establishes  
that the flow curves are essentially located in the 
thermal sector described by our glass/jamming rheology model. 
In particular, the transition from a Newtonian fluid 
to a yield stress solid occurs at low P\'eclet number. 

We have fitted the flow curves to the additive model. 
The fitting parameters are summarized in Table~\ref{tab1}. 
Note that the particle softness $k_BT/\epsilon = 10^{-8}$ 
used in our analysis is somewhat arbitrary, since this parameter 
does not strongly affect the flow curves in this thermal regime,
as expected from the nearly hard sphere nature of the particles.
As shown in Fig.~\ref{fig4}, the transition between fluid and solid
states is very well described by the glass rheology, and the 
fits indicate that the transition occurs for the volume 
fraction $\phig = 0.575$. This is consistent with the analysis performed 
in the original paper~\cite{pete}. This value is also 
consistent with a mode-coupling analysis of the microscopic 
dynamics of colloidal hard spheres, for instance using 
light-scattering~\cite{bill,gio} or microscopy techniques~\cite{weeks}. 
Thus our analysis confirms that the dynamic range probed in a typical 
rheology experiment is not broad enough for the crossover 
to activated dynamics at large density (which has been observed 
numerically~\cite{tom2} and using other
experimental approaches~\cite{gio}) to become apparent.

Although it is clearly glass physics that controls the overall features of the 
flow curves, it is interesting to ask whether jamming 
physics also manifests itself. We argue that in Fig.~\ref{fig4}
there are two distinct aspects where this is the case. 
First, we note that the yield stress reaches the value 
$\sigma^{\rm yield}/\sigma_{\rm T} \approx 10^2$ for 
the largest volume fraction, $\varphi = 0.62$. 
This large value implies that the density is not very far 
from the jamming density so that the divergence in Eq.~(\ref{ys2})
starts to become relevant and a simplified expression for the yield stress, 
such as Eq.~(\ref{ys1}), would not account for the experimental data
taken deep in the glass phase. 

Accordingly, if the density is close enough
to $\phij$, then the athermal Newtonian viscosity $\eta_0(\varphi)$ should 
also start to become large, as it is controlled by a similar 
diverging expression, see Eq.~(\ref{jphi}). 
From the above discussion of the model, see for instance Fig.~\ref{fig3}, 
this viscosity is contributing to the flow curves at large 
P\'eclet number. This is demonstrated in Fig.~\ref{fig4} by the 
dashed line representing the jamming contribution to the total shear 
stress, $\sigma_{\rm J} \approx \eta_0(\varphi) \gammadot$. 
However, while our model predicts that $\eta_0$ grows rapidly when 
$\varphi \to \phij$, such a growth is not observed experimentally, 
and there are clear deviations between the fit and the data for the largest
density in Fig.~\ref{fig4}. This might indicate that the PMMA colloids
cease to behave as nearly hard spheres at large densities and large 
P\'eclet numbers. 

We suggest that it would be interesting 
to repeat such steady state rheological measurements 
with slightly larger PMMA particles
so that large P\'eclet numbers and large densities are more easily studied.
The crossover between glass and jamming 
rheology should then become more apparent and its features 
could be elucidated experimentally in this well-studied 
colloidal system.  

\subsection{Jamming physics: Aqueous foam}

We now analyze the flow curves of aqueous foam reported by 
Herzhaft {\it et al.}~\cite{foam}. 
Foams are considered as prototypical materials displaying 
a jamming transition, because their are typically made of 
non-colloidal soft bubbles. It is worth mentionning that the 
harmonic sphere model considered numerically in Sec.~\ref{harmonic}
was first devised to study the jamming rheology of wet foams~\cite{durian}. 
  
\begin{figure}
\includegraphics[width=\hsize]{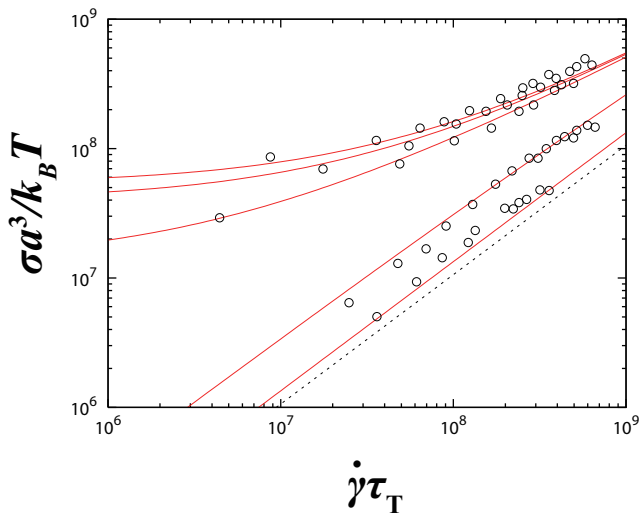} 
\caption{The experimental flow curves of aqueous foam in thermal units, 
after Herzhaft {\it et al.}~\cite{foam}.  
The red curves are fit by the additive model with 
$k_BT/\epsilon = 2.10^{-11}$ and $\phij = 0.55$; see also 
Table~\ref{tab1}. The fluid to solid transition occurs in the jamming 
sector, with negligible influence from thermal fluctuations.} 
\label{fig5}
\end{figure}

The experimental 
system is composed of nitrogen bubbles dispersed in an aqueous 
polymer solution. 
Flow curve were measured at various densities. However, 
the particle size also changes with the density. 
We extracted the 
average particle diameter for each density from the 
reported particle size distributions. 
Note that the typical diameter is rather large at $27 \, \mu{\rm m}$, 
with some size polydispersity. 
The reported solvent viscosity is $\eta_s = 13$~mPa$\cdot$s. 
Using these values, we have estimated 
the thermal time scale and stress scale for each density.  
Typical values are $\tau_{\rm T} = 5 \cdot 10^{5} \, {\rm s}$ 
and $\sigma_{\rm T} = 2 \cdot 10^{-7} \, {\rm Pa}$, 
which are clearly outside the experimentally 
accessible windows for typical time and stress measurements. 
We represent the experimental flow curves using these thermal
units in Fig.~\ref{fig5}. As expected  
this produces large dimensionless numbers 
(such as $\gammadot \tau_{\rm T} \sim 10^7 - 10^9$). Clearly, 
these flow curves belong to the athermal sector, and should 
mainly be controlled by the jamming contribution to the shear stress. 

In the case of a bubble or droplet, we can estimate the particle softness 
using the surface tension $A$.  
The pressure inside a bubble is  larger by $\Delta P = 4A/a$ than on the 
outside, $\Delta P$ being the Laplace pressure. 
This pressure difference acts as a repulsive force between two 
overlapping particles. 
When the overlap length is $d$, the interface area between 
the `overlapping' bubbles is 
given by $\pi a^2 d/2$, 
and the repulsive force becomes $2 \pi A a d$. 
Interpreting this force as the derivative of a pair 
interaction potential, we estimate the softness as 
$\epsilon = 2 \pi A a^2$. 
This expression and the reported surface tension $A$ lead us to estimate 
the dimensionless particle softness for this system 
to be $k_BT/\epsilon = 3 \cdot 10^{-11}$, which is indeed very close to the 
athermal limit in which jamming physics should be observed. 

We fitted the experimental flow curves to our model equation using this 
particle softness. 
The result and fitting parameters are shown in Fig.~\ref{fig5} and 
Table~\ref{tab1}.  
The transition is well fitted as a jamming transition with $\phij = 0.55$. 
Note that this value is much lower than the random close packing density 
usually quoted for spherical particles, $\phij \approx 0.64$. 
The reason of this deviation is not very clear.
We can invoke the fact that the interaction between real bubbles
is more complicated than in simple models of repulsive spheres, 
or the idea that additional microscopic dissipation channels exist
in the real material, 
due for instance to some internal degrees of freedom 
of the bubbles.
We note, however, that in other studies the jamming transition 
of a real foam was identified and located very close to the 
jamming transition occurring in simple harmonic sphere 
models, both at the level of static~\cite{martin} and 
rheological~\cite{stjalmes} properties.   

\subsection{Exploring the crossover: Oil-in-water emulsions}

\label{emulsion}

\begin{figure}
\includegraphics[width=\hsize]{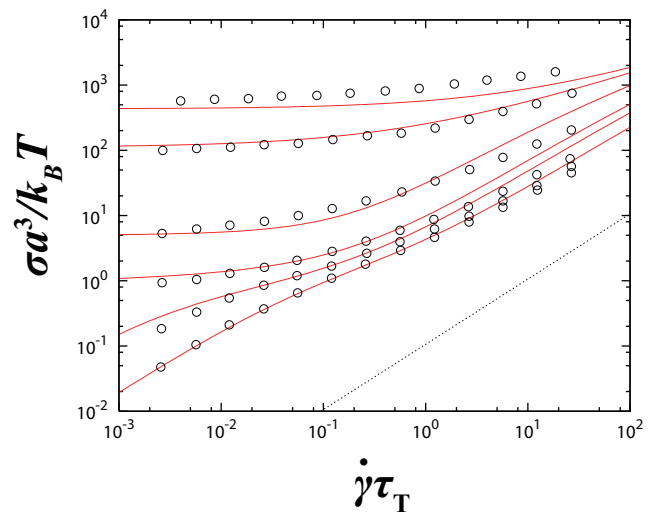} 
\caption{Experimental flow curves of oil-in-water 
emulsion in thermal units, after Mason {\it et al.}~\cite{bibette}.
The red curves are fits to the additive model with $k_BT/\epsilon = 
3\cdot 10^{-7}$, $\phig = 0.579$ and $\phij = 0.62$; see also Table~\ref{tab1}.
While a glass transition is observed at low shear rates, 
the sharp increase of the yield stress in the glass phase 
is a signature of the jamming transition.} 
\label{fig6}
\end{figure}

We next focus on the flow curves obtained for 
oil-in-water emulsions, reported by Mason {\it et al.}~\cite{bibette}.  
The system is composed of oil droplets 
stabilized with sodium dodecyl sulfate, which are dispersed in water.
Although experiments with different particle diameters were 
performed~\cite{bibette}, 
we only analyze the results for droplet diameter 
$a = 0.5 \, \mu{\rm m}$ (with less than 10~\% size dispersity), 
because the most complete set of flow curves were reported for this 
specific particle size~\cite{bibette}. 
Using the water viscosity value of $\eta_s = 1$~mPa$\cdot$s, 
we estimate the thermal time and 
stress scales to be $\tau_{\rm T} = 0.25 \, {\rm s}$ 
and $\sigma_{\rm T} = 3 \cdot 10^{-2} \, {\rm Pa}$, respectively. 
These values are well within the measurable range and then we expect 
aspects of the glass transition to be relevant, although 
large P\'eclet numbers can certainly be accessed too. 
However, 
it is important to note
that, in contrast to
PMMA colloids as discussed in Sec.~\ref{pmma}, emulsions 
are made of soft droplets which can therefore easily be compressed above the 
jamming (or `random close packing') density. This implies 
that aspects of the jamming transition are also 
potentially important for this system. The fact 
that features of both transitions are relevant 
for these emulsions is already apparent from the original experimental 
papers: the linear rheology of these systems was interpreted 
using the concepts of $\alpha$ and $\beta$ relaxations 
and power law scalings near the glass transition density $\phig$
inspired by mode-coupling theory~\cite{mason},
while the density dependence of the yield stress was fitted to a power 
law near the jamming density $\phij$ in a separate 
article~\cite{bibette}. Our additive model represents an ideal 
framework to unify and rationalize these findings.

The complex rheological features summarized above can be seen 
in Fig.~\ref{fig6}, where the 
experimental flow curves are shown in thermal units. The system
behaves as a Newtonian fluid at low enough density, $\varphi \leq 0.57$, 
and as a yield stress fluid for larger densities. However, the behaviour 
of the yield stress in the glass phase is clearly nontrivial, as it increases 
sharply with $\varphi$ in the range $\varphi \approx 0.60 - 0.62$ 
to reach large dimensionless values, 
$\sigma^{\rm yield} / \sigma_{\rm T} \approx 4 \cdot 10^2$ for $\varphi = 0.65$. 
This final shear stress value should belong to the 
athermal regime, possibly suggesting a crossover between 
the glass and jammed phases.

To confirm these qualitative conclusions, we 
fit the experimental flow curves using the additive rheological model.
To this end, we first estimate the particle softness from 
the surface tension reported in the experimental article~\cite{bibette},
using the method outlined in the previous subsection for aqueous foams. 
We obtain $k_BT/\epsilon = 3 \cdot 10^{-7}$. Interestingly, this value 
is much larger than the one obtained for both foams and PMMA colloids, 
but remains at the lower end of the range of temperatures simulated 
in our numerical simulations~\cite{letter}; see Fig.~\ref{fig1}. 
We then fitted the experimental flow curves to the model equation using this 
softness value, and obtained the fits shown with lines 
in Fig.~\ref{fig6}. The corresponding fitting parameters are
listed in Table.~\ref{tab1}. We obtain 
the glass and jamming transition densities at
$\phig = 0.579$ and $\phij = 0.62$, these values being consistent with 
previous experimental analysis~\cite{mason,bibette}. 
We emphasize that both the fluid to solid transition observed at 
low P\'eclet numbers, and the sharp increase with density of the yield stress 
are well fitted by our model. This allows us to conclude unambiguously  
that the transition which appears near $\varphi = 0.58$ 
is a colloidal glass transition, while that the sharp stress increase 
at higher density near $\varphi \sim 0.62$ 
is a clear signature of the jamming transition. 
We stress that to obtain a quantitative description of these flow curves, 
it is absolutely necessary that both glass and jamming contributions to
the shear stress are combined in a unified model as in 
Eq.~(\ref{full}). Therefore, the 
excellent fits in Fig.~\ref{fig6} provide strong experimental support 
for the additive model and theoretical analysis 
offered in the present work. As a direct physical 
corollary, we also conclude that 
the shear rheology of micron-size emulsions is quite complex, 
as the flow curves reveal all the features described 
by the most complex version of our additive rheological model.

\begin{figure}
\includegraphics[width=\hsize]{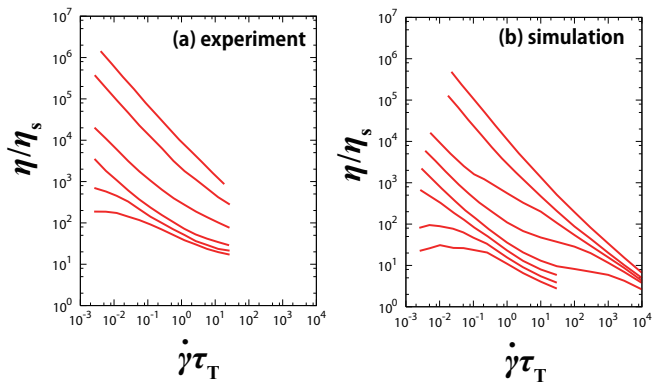} 
\caption{Viscosity vs.\ shear flow curves measured 
(a) experimentally for 
oil-in-water emulsions~\cite{bibette} and (b)
numerically for harmonic spheres at $k_BT /\epsilon = 10^{-6}$.
We use similar dimensionless scales to emphasize the similarity
between the two situations, suggesting that emulsions are an ideal
experimental system to study the glass-jamming crossover.} 
\label{fig7}
\end{figure}

As shown in Table~\ref{tab1}, the parameters used to describe the emulsion 
data are very similar to the ones used to fit the flow curves 
of harmonic spheres in Fig.~\ref{fig1}. This implies that 
the two sets of data should in fact be quantitatively very close. 
To show this more directly, we replot the flow curves 
for both these systems  
in a slightly different representation in Fig.~\ref{fig7}, where 
the shear rate dependence of the effective viscosity,  
$\eta(\gammadot) \equiv \sigma(\gammadot)/\gammadot$, 
is shown. We make the experimental viscosity dimensionless 
using the solvent viscosity $\eta_s$, and combine 
Stokes' law in Eq.~(\ref{stokes}) with Eq.~(\ref{taut}) to obtain 
the bare viscosity
$\eta_s = \xi /(3 \pi a)$ in the numerics.
The results in Fig.~\ref{fig7} 
are clearly very similar. This means that, somewhat surprisingly, 
modelling of real 
emulsions as harmonic spheres with overdamped Langevin 
dynamics is quantitatively very accurate. 
Our results also suggest that an interesting athermal Newtonian behavior
characterized by the viscosity $\eta_0$  
would appear experimentally if higher 
shear rates could be studied. (Note that, unusually, the range of 
shear rates studied
is actually broader here in the numerical work than in the experiments).
This second viscosity should be directly
controlled by the jamming transition, and emulsions could thus be used to 
improve our understanding of its density dependence, which is usually studied  
using non-Brownian granular suspensions~\cite{lindner,boyer}.
We suggest that a more extensive rheological investigation of emulsions, 
perhaps by varying the droplet size in a more systematic manner to 
gradually access larger P\'eclet numbers, would be very interesting. 
We have found literature data for much larger particle 
diameters~\cite{otsubo,otsubo2}, but the density range covered in these  
flow curves was too limited to perform a detailed analysis 
of their evolution using our model.

\begin{table*}
\small
\caption{Summary of fitting parameters used in the paper. 
For emulsion, we used the same values of $Y_{\rm J}$ and 
$\beta_{\rm J}$ as in the original article~\cite{bibette}.}
\label{tab1}
\begin{tabular*}{\textwidth}{@{\extracolsep{\fill}}llllllllllllllllll}
\hline
System & $p_{\rm G}$ & $p_{\rm J}$ & $h_{\rm G}$ & $h_{\rm J}$ & $\alpha_{\rm G}$ & $\alpha_{\rm J}$ & $\gamma_{\rm G}$ & $\gamma_{\rm J}$ & $Y_{\rm G}$ & $Y_{\rm G}'$ & $Y_{\rm GJ}$ & $Y_{\rm J}$ & $\beta_{\rm G}$ & $\beta_{\rm GJ}$ & $\beta_{\rm J}$  \\
\hline
Simulation~\cite{letter} & 7 & 4 & 0.03 & 0.1 & 0.3 & 0.5 & 2.2 & 2.0 & 0.15 & 0.25 & 0.02 & 0.02 & 0.6  & 1.0 & 1.0 \\
PMMA colloids~\cite{pete} & 3 & 4 & 0.03 & 0.3 & 0.3 & 0.5 & 2.2 & 2.0 & 1.5 & 6.0 & 0.02 & 0.02 & 0.6  & 1.0 & 1.0 \\
Foam~\cite{foam} & 7 & 12 & 0.03 & 0.08 & 0.3 & 0.6 & 2.2 & 2.0 & 0.15 & 0.25 & 0.02 & 0.01 & 0.6  & 1.0 & 1.0 \\
Emulsion~\cite{bibette} & 7 & 0.5 & 0.03 & 0.06 & 0.3 & 0.5 & 2.2 & 2.0 & 0.25 & 0.65 & 0.02 & 0.016 & 0.6 & 1.0 & 2.0 \\
PNIPAM~\cite{carrier} &  7 & 1 & 0.03 & 0.07 & 0.5 & 0.4 & 2.2 & 2.0 & 12 & 30 & 1.5 & 0.03 & 0.6  & 1.0 & 1.0 \\
PNIPAM~\cite{yodh} & 30 & 1 & 0.2 & 0.005 & 0.5 & 0.4 & 2.2 & 2.0 & 15 & 500 & 3.5 & 3 & 0.6  & 1.0 & 1.0 \\
\hline
\end{tabular*}
\end{table*}

\subsection{PNIPAM microgels: Glass or jamming?}

In this final experimental subsection, 
we analyze the steady state flow curves obtained for two 
independent sets of similar PNIPAM microgels. These particles 
are currently the focus of a large number 
of investigations~\cite{microgelbook}, 
for at least two reasons.
First, these particles are made of microgels, and are therefore 
very soft. As such, they represent a new type of colloidal 
system, potentially very distinct from the more heavily
studied hard sphere paradigm~\cite{microgelnature}. 
A second important feature is that
PNIPAM microgels are highly sensitive to temperature, in the sense 
that a small temperature change induces a relatively large change of 
the particle size. This endows microgel suspensions with 
peculiar physical properties~\cite{microgelbook}. 

We have analyzed two sets of flow curves obtained experimentally.
One is reported by Carrier {\it et al.}~\cite{carrier}
and was interpreted there in the framework of the mode-coupling 
theory extended to sheared fluids~\cite{fuchs}, thus implying the 
assumption that 
the fluid to solid transition in microgel suspensions
is a colloidal glass transition.  
The second one is reported by Nordstrom {\it et al.}~\cite{yodh}.
In this article, the flow curves are scaled using 
a procedure that was first employed in the context of the
jamming transition of soft spheres~\cite{olson}, and the authors 
interpret the fluid to solid transition as the result
of jamming. Our model, based as it is on a careful distinction 
between glass and jamming 
physics, is thus
a natural tool to resolve the conflict between two opposing 
interpretations of the data.  

\begin{figure}
\includegraphics[width=\hsize]{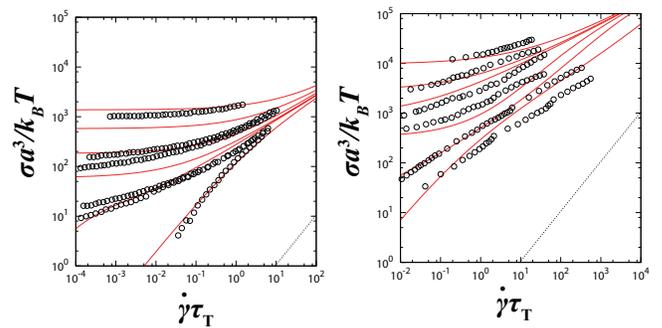} 
\caption{Experimental flow curves of PNIPAM microgels of 
(left) Carrier {\it et al.}~\cite{carrier} 
and (right) Nordstrom {\it et al.}~\cite{yodh}, using
thermal units. The red curves are fit by the additive model with 
$k_BT/\epsilon = 10^{-4}$
in both cases; see also Table~\ref{tab1}.  
The corresponding glass transition densities 
are $\phig = 0.68$ (left) and $\phig = 0.638$ (right).}
\label{fig8}
\end{figure}

Like for other systems, 
our first task is to replot the flow curves in dimensionless 
units, using the particle size and solvent viscosity to determine 
the relevant thermal units; see Fig.~\ref{fig8}.
In the first experiment~\cite{carrier}, 
the authors had reported their flow curves using similar thermal units. 
However, the particle radius was used as a microscopic length scale.
To be consistent with the analysis performed elsewhere in the paper, 
we replotted the results using the particle diameter as the 
unit length. In the second experiment~\cite{yodh}, 
the results are reported in units of ${\rm Pa}$ 
and ${\rm s^{-1}}$ for shear stress and shear rate, respectively. 
The authors also reported the particle diameter as a function of temperature, 
and the corresponding packing 
fraction~\footnote{In PNIPAM, the particle size, and thus the
packing fraction, are controlled by temperature. 
Note that this temperature change is very small, 
and then, its direct effect on the particle softness $k_BT/\epsilon$ 
can be neglected.}.
Using these diameters and the reported solvent viscosity, 
$\eta_s = 1 \, {\rm mPa}\cdot{\rm s}$, 
we converted these flow curves to the dimensionless 
representation shown in Fig.~\ref{fig8}. Once this rescaling is performed, 
it becomes evident that the experimental results 
obtained by the two research groups extend over {\it a very similar range 
of shear stresses and shear rates}. 
Thus the simple dimensional procedure indicates that the 
physics probed in these two experiments is the same.
In particular, we notice immediately that the experiments
are performed over a range of shear rates corresponding 
to relatively small P\'eclet numbers, down to 
$P_{\rm e} = 10^{-4}$. This is because the particle size is small enough 
for thermal fluctuations to be relevant over the typical timescale
of the experiments. 
Note that when converted into dimensionless units using the athermal
timescale $\tau_0$, as done in Ref.~\cite{yodh}, the typical time scales
extracted from the flow curves 
become unphysically large, $\tau / \tau_0 \sim 10^7 - 10^{11}$, 
again indicating that the experimental data 
do not belong to the athermal jamming sector. 

An independent confirmation of the relevance of thermal fluctuations 
for PNIPAM microgels is obtained from the linear viscoelastic experiments
performed by Carrier {\it et al.}~\cite{carrier}. 
Their results show that the loss modulus $G''(\omega)$ becomes larger than the 
storage modulus $G'(\omega)$ at low frequency. 
This indicates that spontaneous relaxation induced by thermal 
motion plays an important role in the rheology. Qualitatively similar 
results~\cite{mason} have been obtained, for instance, for the 
emulsion system discussed 
in Sec.~\ref{emulsion}. 

In order to fit the experimental data on PNIPAM microgels to our 
additive model, we have to estimate 
the particle softness $k_BT/\epsilon$. 
However, it is not obvious that the interaction between 
microgel particles can be approximated with a soft harmonic repulsion.
Fortunately, a quantitative test of this assumption was performed
in a detailed analysis of the vibrational properties of 
the amorphous solid phase of PNIPAM particles~\cite{chen}.
For a similar system this study establishes the quality of the mapping 
and obtains $k_BT / \epsilon = 2 \cdot 10^{-5}$. A slightly larger 
value, near $k_BT / \epsilon = 10^{-4}$,
was later determined through a theoretical 
analysis of the mean-squared displacements of PNIPAM 
particles~\cite{ikeda}.

Using $k_BT / \epsilon = 10^{-4}$ for the
value of the particle softness,
we then fitted both sets of 
flow curves to our additive model. 
The result of this procedure is shown in Fig.~\ref{fig8}. The quality of 
the fits is similar to the one obtained within the mode-coupling theory
framework~\cite{carrier}. As is clear also from the numerical flow curves 
obtained for a similar temperature in Fig.~\ref{fig1}, jamming physics
plays virtually no role in these fits, and the main 
control parameter to be adjusted is the glass transition density, 
which we estimate as $\phig = 0.68$ and $\phig = 0.638$ for 
the first~\cite{carrier} and second~\cite{yodh} set of experiments, 
respectively.
Therefore, we conclude that the transition between a viscous fluid 
and a soft solid observed at low shear rate in the data of 
Fig.~\ref{fig8} is a {\it colloidal glass transition}, thus favoring 
the interpretation of Carrier {\it et al.}~\cite{carrier} 
over the one of Nordstrom {\it et al.}~\cite{yodh}.

We notice that another effect of the large particle softness 
is that the glass transition density does not correspond to 
the one of the hard sphere system discussed in Sec.~\ref{pmma}, but 
is considerably larger. 
This results from the fact that at finite temperatures 
microgel particles can overlap while hard spheres cannot, thereby
shifting the glass transition density to larger values, 
as observed numerically~\cite{tom,tom2} and captured theoretically 
using a mode-coupling approach~\cite{szamel}. 
We emphasize that the quantitative 
similarity between the glass transition density 
of harmonic spheres and the jamming transition of hard sphere
particles is then nothing but a numerical coincidence. 

Another difference between hard and soft particles lies in the value
of the typical yield stress scales observed in the experiments, see
Fig.~\ref{fig8}, which is somewhat larger for the soft 
microgel particles. To account for this feature, we had to 
use a larger value of the parameter $Y_G'$ (see Table~\ref{tab1})
which sets the scale of the yield stress in Eq.~(\ref{yg}).
We note that a similar change of numerical prefactor by about 
an order of magnitude was introduced (with no discussion) 
in the mode-coupling analysis of Carrier {\it et al.}~\cite{carrier}.
We speculate that the effect could be due to the influence of polymeric 
internal degrees of freedom of the microgel particles. 

While the microgel rheology analyzed in this subsection 
leads us to the conclusion that dense PNIPAM assemblies should 
be considered as thermal glasses rather than athermal jammed solids, 
we suggest that performing experiments with larger PNIPAM 
particles would be extremely rewarding, as this would produce 
an experimental system where large P\'eclet numbers could be studied
with particles that are very soft. This would allow the 
jamming transition and rheology of soft particle systems to be studied 
experimentally~\cite{martin,xchen,sessoms}. 

\section{Summary and perspectives}

\label{conclusion}

In this paper, we described in detail 
the `additive' rheological model first introduced 
in our numerical analysis of the interplay between 
glass and jamming rheologies~\cite{letter}, and we used 
the model to revisit published experimental data.

In our model,  the stress contributions 
from thermal glass and athermal jamming physics are 
treated separately and we assume they can be linearly 
combined in a unified model. Because we want to 
describe the emergence of amorphous solids at large density
for both colloidal and non-Brownian repulsive particles,  
the main control parameters for the formation of soft solids 
are therefore the particle packing fraction, $\varphi$, 
and the particle softness expressed in units of thermal 
energy, $k_BT/\epsilon$. From the analysis of numerical
data obtained for harmonic spheres at both finite and 
zero temperatures, we developed simple functional forms 
for the flow curves $\sigma(\gammadot)$ measured at a given
state point, for the Newtonian viscosities $\eta_T$ and $\eta_0$ 
(in the fluid regimes), and for the yield stress
$\sigma^{\rm yield}$ (in the solid phases). 

\begin{figure}
\includegraphics[width=\hsize]{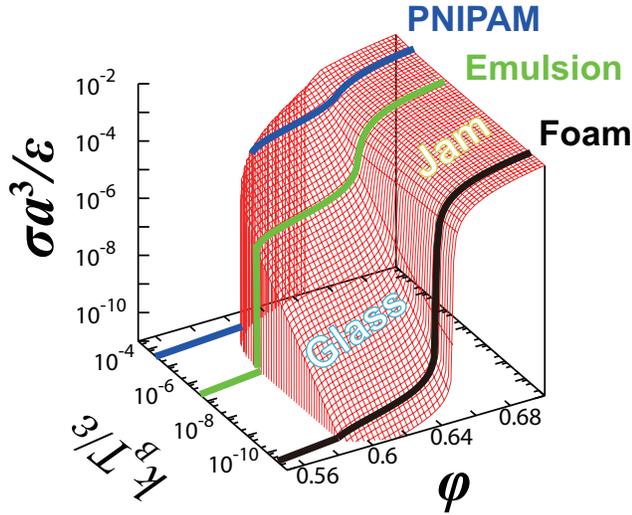} 
\caption{Three-dimensional `jamming phase diagram' showing 
the yield stress surface as a function of the thermodynamic
parameters temperatures and density, in a dimensionless 
representation (particle softness $k_BT/\epsilon$, volume fraction 
$\varphi$, and stress $\sigma a^3/\epsilon$).
The three lines represent the location of the experimental 
systems discussed in 
Sec.~\ref{experiments}. Foams are mainly sensitive to jamming 
physics, emulsions display an interesting interplay between 
glass and jamming transitions, while PNIPAM microgels 
undergo a colloidal glass transition.}
\label{fig9}
\end{figure}

In particular, our additive model describes 
the emergence of solidity for soft repulsive systems 
as a function of temperature and volume fraction. As a result, 
we can use the yield stress fits shown in 
Fig.~\ref{fig2} to construct the three-dimensional `jamming 
phase diagram'~\cite{liunagel,pica} shown in Fig.~\ref{fig9}, which represents 
the yield surface delimiting the solid and fluid phases 
for all values of the thermodynamic control parameters~\cite{letter}.
This surface shows that the emergence of solidity at finite 
temperatures is always controlled by the colloidal glass transition,  
the $T=0$ point being the only situation where the jamming 
transition signals also the onset of solidity.  

The line describing the PNIPAM microgel 
data in Fig.~\ref{fig9} corresponds to high temperatures 
(soft particles), and therefore the jamming transition does not
influence the rheology of these microgel suspensions. 
When temperature is decreased, a clearer signature 
of the jamming transition is observed as a sharp increase of the 
yield stress with density. In this regime, the jamming transition
appears as a change in the nature of the glass phase~\cite{PZ,hugo}, 
but it does 
not control the emergence of solidity which still occurs
at the glass transition. This description applies to emulsions, 
as shown in Fig.~\ref{fig9}. 
Finally, when thermal effects become negligible, the thermal yield 
stress might become too small to be detected experimentally, and solidity
genuinely emerges at the jamming transition. This is the case for 
foams in Fig.~\ref{fig9} for which the glass `wing' has negligible 
effects. Note that PMMA colloidal suspensions would appear at nearly the 
same temperature/softness as foams in the jamming phase diagram of 
Fig.~\ref{fig9}. However, with the particle size being much smaller than 
for foams, the yield stress emerging at the colloidal glass 
transition would easily be measured experimentally, and the measurements
would stop as the jamming density is approached because the yield
stress would seem to diverge there. 

\begin{figure}
\includegraphics[width=\hsize]{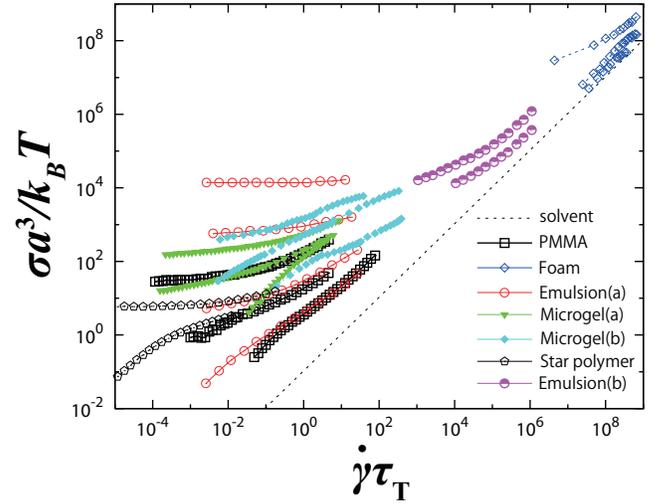} 
\caption{Superposition of experimental flow curves for different
materials using thermal units. 
PMMA colloids with $a = 0.36 \, {\rm \mu m}$, after Petekidis {\it 
et al.}~\cite{pete}. 
Foam with $a=27 \, {\rm \mu m}$, after Herzhaft {\it et al.}~\cite{foam}. 
Emulsion(a) with $a=0.5 \, {\rm \mu m}$, after Mason 
{\it et al.}~\cite{bibette}. 
PNIPAM(a) with  $a=0.2 \, {\rm \mu m}$, after 
Carrier {\it et al.}~\cite{yodh}. 
PNIPAM(b) with $a=1.2 \, {\rm \mu m}$, after Nordstrom 
{\it et al.}~\cite{yodh}. 
Star polymers with $a=0.07 \, {\rm \mu m}$, after 
Koumakis {\it et al.}~\cite{star}. 
Emulsion(b) with $a=8 \, {\rm \mu m}$, after 
Otsubo {\it et al.}~\cite{otsubo}.}
\label{fig10}
\end{figure}

As shown by the jamming phase diagram in Fig.~\ref{fig9}, 
our analysis is useful 
in organizing the physics of different experimental systems.  
To confirm this, we have used our  
additive rheological model to analyze various experimental flow curves
obtained for a variety of dense suspensions. 
The systems we focussed on were PMMA colloids~\cite{pete}, 
aqueous foam~\cite{foam}, oil-in-water emulsions~\cite{bibette}, 
and PNIPAM microgels~\cite{carrier,yodh}. 
We have also gathered experimental data 
from other sources, in particular ultrasoft particles 
composed of star polymers~\cite{star}, 
and data for emulsions with larger droplet sizes~\cite{otsubo}, 
but for brevity the  results of our analysis have not been presented in 
Sec.~\ref{experiments}. 

We showed that all the above experimental results can be 
successfully analyzed using the additive model. It 
is instructive to replot all data in a single figure
using the dimensional procedure adopted throughout this 
paper, i.e.\ expressing stress and time scales in thermal 
units $\sigma_{\rm T}$ and $\tau_{\rm T}$, see Eqs.~(\ref{taut}, 
\ref{sigmat}). These flow curves are collected in 
Fig.~\ref{fig10}.
In this representation, the flow curves for PMMA colloids, 
star polymers, PNIPAM microgels lie in the same
sector, which corresponds to the thermal sector in our model; 
see Fig.~\ref{fig3}. Therefore, the formation of amorphous 
solids in these systems stems from the physics of 
the colloidal glass transition. On the other hand, 
foams lie outside this regime and are controlled, accordingly,  
by the jamming transition. Interestingly, emulsions 
lie somewhat in between and so are  influenced by 
both types of physics, as discussed in Sec.~\ref{emulsion}.  
Note in particular that emulsions with larger 
droplet sizes, also  shown in Fig.~\ref{fig10}, 
could be useful systems to fill the gap between colloids and foams.
While experimental studies of microgel particles
have been interpreted from the point of view of the jamming 
transition~\cite{yodh}, 
our analysis shows that for these soft colloidal 
particles the physics of jamming has, in fact, only a negligible effect. 
A similar conclusion has recently been reached based on the analysis
of the short-time vibrational dynamics in the amorphous 
phase~\cite{ikeda}.

Our conclusion that glass and jamming rheologies belong to different 
sectors and contribute linearly to the shear stress is
directly supported by the numerical flow curves obtained for harmonic
spheres, and by the analysis of the oil-in-water emulsions 
in Sec.~\ref{emulsion} which clearly showed the complex features also  
observed in the simulations. We mentioned that similar
indications are also found for PMMA colloids, in particular 
at large P\'eclet number and larger density, while microgel
suspensions appear less well suited for a detailed experimental
investigations of the interplay between glass and jamming rheology.
The data in Fig.~\ref{fig10} show that many current 
experimental systems are in fact too close to the thermal
sector. Therefore, to fill the gap between small colloids and 
foams, measurements on sufficiently hard particles (PMMA colloids, 
emulsions) with particle sizes in the range $a=0.5-5 \, \mu{\rm m}$ 
would appear ideally suited. We hope our work will stimulate experimental
investigations in this direction.

\acknowledgments

We thank M. Cloitre, Y. Otsubo, G. Petekidis, and S. Teitel 
for useful correspondence.
We thank R\'egion Languedoc-Roussillon (L. B., A. I.) and 
JSPS Postdoctoral Fellowship for Research Abroad (A. I.) for financial 
support. 
The research leading to these results has received funding
from the European Research Council under the European Union's Seventh
Framework Programme (FP7/2007-2013) / ERC Grant agreement No 306845.


\begin{thebibliography}{99}

\bibitem{larson}
R. G. Larson, {\it The Structure and Rheology of Complex Fluids} (Oxford
University Press, New York, 1999).

\bibitem{coussot}
P. Coussot, {\it Rheometry of Pastes, Suspensions, and Granular
Materials} (Wiley, New York, 2005).

\bibitem{rmp}
L. Berthier and G. Biroli,
Rev. Mod. Phys. {\bf 83}, 587 (2011).

\bibitem{jammingrev}
A. J. Liu and S. R. Nagel, 
Annual Reviews of Cond. Mat. Phys. {\bf 1}, 347 (2010).

\bibitem{pusey}
P. N. Pusey and W. van Megen, Nature (London) {\bf 320}, 340
(1986).

\bibitem{liunagel}
A. J. Liu and S. R. Nagel, Nature {\bf 396}, 21 (1998).

\bibitem{grains}
{\it Jamming and rheology},
Eds.: A. J. Liu and S. R. Nagel (Taylor and Francis,
New York, 2001).

\bibitem{ohern1}
C. S. O'Hern, S. A. Langer, A. J. Liu, and S. R. Nagel,
Phys. Rev. Lett. {\bf 88}, 075507 (2002).

\bibitem{vanhecke} M. van Hecke, J. Phys.: Condens. Matter 
{\bf 22} 033101 (2010).

\bibitem{book} 
{\it Dynamical heterogeneities in glasses, colloids and
granular materials}, Eds.: L. Berthier, G. Biroli, J.-P. Bouchaud,
L. Cipelletti, and W. van Saarloos (Oxford University Press, Oxford, 2011).

\bibitem{yamamoto}
R. Yamamoto and A. Onuki, Phys. Rev. E {\bf 58}, 3515 (1998).

\bibitem{BBK} L. Berthier, J.-L. Barrat, and J. Kurchan
Phys. Rev. E {\bf 61}, 5464 (2000).

\bibitem{BB}
L. Berthier and J.-L. Barrat, J. Chem. Phys. {\bf 116}, 6228 (2002).

\bibitem{fuchs}
M. Fuchs and M. E. Cates, Phys. Rev. Lett. {\bf 89}, 248304 (2003).

\bibitem{exp1}
M. Siebenbuerger, M. Fuchs, H. Winter, and M. Ballauff,
J. Rheol {\bf 53}, 707 (2009). 

\bibitem{olson}
P. Olsson and S. Teitel, Phys. Rev. Lett. {\bf 99}, 178001 (2007).

\bibitem{tighe}
B. P. Tighe, E. Woldhuis, J. J. C. Remmers, W. van Saarloos, and 
M. van Hecke, Phys. Rev. Lett. {\bf 105}, 088303 (2010). 

\bibitem{olson2}
P. Olsson and S. Teitel,
Phys. Rev. Lett. {\bf 109}, 108001 (2012).

\bibitem{lerner} 
E. Lerner, G. D\"uring, and M. Wyart, 
Proc. Natl. Acad. Sci. USA {\bf 109}, 4798 (2012).

\bibitem{egami} P. Guan, M. Chen, and T. Egami,
Phys. Rev. Lett. {\bf 104}, 205701 (2010).  

\bibitem{yodh} K. N. Nordstrom, E. Verneuil, P. E. Arratia, 
A. Basu, Z. Zhang, A. G. Yodh, J. P. Gollub, and D. J. Durian,
Phys. Rev. Lett. {\bf 105}, 175701 (2010).  

\bibitem{letter}
A. Ikeda, L. Berthier, and P. Sollich,
Phys. Rev. Lett. {\bf 109}, 018301 (2012).

\bibitem{tom}
L. Berthier and T. A. Witten,
EPL {\bf 86}, 10001 (2009). 

\bibitem{durian} D. J. Durian, Phys. Rev. Lett. \textbf{75}, 4780 (1995);
Phys. Rev. E {\bf 55}, 1739 (1997). 

\bibitem{allen}
M. Allen and D. Tildesley, {\it Computer Simulation of Liquids}
(Oxford University Press, Oxford, 1987).

\bibitem{olson3}
P. Olsson and S. Teitel, arxiv:1211.2839.

\bibitem{thomas}
T. Voigtmann, Eur. Phys. J. E {\bf 34}, 106 (2011).

\bibitem{chaikin}
Z. Cheng, J. Zhu, P. M. Chaikin, S. E. Phan, and W. B.
Russel, Phys. Rev. E {\bf 65}, 041405 (2002).

\bibitem{bill}
W. van Megen, T. C. Mortensen, S. R. Williams, and J. Muller, 
Phys. Rev. E {\bf 58}, 6073 (1998).

\bibitem{gio}
G. Brambilla, D. El Masri, M. Pierno, L. Berthier, 
L. Cipelletti, G. Petekidis, and A. B. Schofield, 
Phys. Rev. Lett. {\bf 102}, 085703 (2009). 

\bibitem{gotze}
W. G\"otze, {\it Complex Dynamics of Glass-Forming Liquids:
A Mode-Coupling Theory} (Oxford University Press, Oxford,
2008).

\bibitem{tom2}
L. Berthier and T. A. Witten,
Phys. Rev. E {\bf 80}, 021502 (2009).

\bibitem{lindner}
C. Bonnoit, T. Darnige, E. Clement, and A. Lindner,
J. Rheol. {\bf 54}, 65 (2010).

\bibitem{boyer}
F. Boyer, E. Guazzelli, and O. Pouliquen,
Phys. Rev. Lett. {\bf 107}, 188301 (2011).

\bibitem{donev}
A. Donev, S. Torquato, F. H. Stillinger, and R. Connelly, 
Phys. Rev. E {\bf 70}, 043301 (2004).

\bibitem{pinaki}
 P. Chaudhuri, L. Berthier, and S. Sastry, Phys. Rev. Lett. {\bf 104}, 165701
(2010).

\bibitem{bookmewis}
J. Mewis and N. J. Wagner, {\it Colloidal suspension rheology}
(Cambridge University Press, Cambridge, 2012). 

\bibitem{freevol}
M. H. Cohen and D. Turnbull, J. Chem. Phys. {\bf 31}, 1164 (1959).

\bibitem{wood}
Z. W. Salsburg and W. W. Wood, J. Chem. Phys. {\bf 37}, 798 (1962).

\bibitem{hugo2}
L. Berthier, H. Jacquin, and F. Zamponi,
Phys. Rev. E {\bf 84}, 051103 (2011).

\bibitem{pete} 
G. Petekidis, D. Vlassopoulos, and P. N.  Pusey, 
J. Phys.: Condens. Matter {\bf 16}, S3955 (2004).

\bibitem{hansen}
J. P. Hansen and I. R. McDonald, {\it Theory of Simple Liquids} (Elsevier,
Amsterdam, 1986).

\bibitem{weeks}
E. Weeks, J. C. Crocker, A. C. Levitt, A. Schofield,
and D. A. Weitz, Science {\bf 287}, 627 (2000).

\bibitem{foam} B. Hertzhaft, S. Kakadjian, and M. Moan,
Colloids Surf. A {\bf 263}, 153 (2005). 

\bibitem{martin}
G. Katgert and M. van Hecke, EPL {\bf 92}, 34002 (2010). 

\bibitem{stjalmes}
A. Saint-Jalmes and D. J. Durian, J. Rheol. {\bf 43}, 1411 (1999).

\bibitem{bibette}
T. G. Mason, J. Bibette, D. A. Weitz, 
J. Colloid Interface Sci. {\bf 179}, 439 (1996).

\bibitem{mason}
T. G. Mason and D. A. Weitz, 
Phys. Rev. Lett. {\bf 75}, 2770 (1995).

\bibitem{otsubo} Y. Otsubo and R. K. Prud'homme,
Rheol. Acta {\bf 33}, 29 (1994). 

\bibitem{otsubo2} Y. Otsubo and R. K. Prud'homme, 
J. Soc. Rheol. Japan {\bf 20}, 125 (1992). 

\bibitem{microgelbook} {\it Microgel suspensions}, 
Eqs.: A. Fernandez-Nieves, H. Wyss, J. Mattsson, D. A. Weitz
(Wiley, Weiheim, 2011).

\bibitem{microgelnature}
J. Mattsson, H. M. Wyss, A. Fernandez-Nieves, K. Miyazaki,
Z. Hu, D. R. Reichman, and D. A. Weitz, 
Nature {\bf 462}, 83 (2009).

\bibitem{carrier} 
V. Carrier and G. Petekidis, J. Rheol. {\bf 53}, 245 (2009). 

\bibitem{chen} 
K. Chen, W. G. Ellenbroek, Z. Zhang, D. T. N. Chen, P. J. Yunker, C. Brito,
O. Dauchot, S. Henkes, W. van Saarloos, A. J. Liu, and A. G. Yodh, Phys.
Rev. Lett. {\bf 105}, 025501 (2010).

\bibitem{ikeda} A. Ikeda, L. Berthier, and G. Biroli
J. Chem. Phys. {\bf 138}, 12A507 (2013).

\bibitem{szamel}  
L. Berthier, E. Flenner, H. Jacquin, and G. Szamel,
Phys. Rev. E {\bf 81}, 031505 (2010).

\bibitem{xchen}
X. Cheng, Phys. Rev. E {\bf 81}, 031301 (2010).

\bibitem{sessoms}
D. A. Sessoms, I. Bischofberger, L. Cipelletti, and V. Trappe,
Philos. Trans. R. Soc. London, Ser. A {\bf 367}, 5013 (2009).

\bibitem{pica}
M. Pica Ciamarra, M. Nicodemi and A. Coniglio, 
Soft Matter {\bf 6}, 2871 (2010).

\bibitem{PZ}
G. Parisi and F. Zamponi, Rev. Mod. Phys. {\bf 82}, 789 (2010).

\bibitem{hugo}
H. Jacquin, L. Berthier, and F. Zamponi, 
Phys. Rev. Lett. {\bf 106}, 135702 (2011).

\bibitem{star}
N. Koumakis, A. Pamvouxoglou, A. Poulos, and G. Petekidis, 
Soft Matter {\bf 8}, 4271 (2012).

\end{thebibliography}
\end{document}